\documentclass[prd,aps,showpacs,nofootinbib,tightenlines]{revtex4}
\usepackage{mathrsfs}
\usepackage{amsmath}
\usepackage{amssymb}
\usepackage{epsfig}
\usepackage{graphicx}
\usepackage{booktabs}
\usepackage{multirow}
\usepackage{subfigure}
\usepackage[section]{placeins}
\usepackage{float}
\usepackage{slashed}
\begin{document}
\newcommand{\psl}{ p \hspace{-1.8truemm}/ }
\newcommand{\nsl}{ n \hspace{-2.2truemm}/ }
\newcommand{\vsl}{ v \hspace{-2.2truemm}/ }
\newcommand{\epsl}{\epsilon \hspace{-1.8truemm}/\,  }

\title{ Estimates of the isospin-violating  $\Lambda_b\rightarrow \Sigma^0 \phi, \Sigma^0 J/\psi$  decays and the $\Sigma-\Lambda$  mixing}
\author{Zhou Rui$^1$}\email[Corresponding  author: ]{jindui1127@126.com}
\author{Jia-Ming Li$^1$}
\author{Chao-Qi Zhang$^1$}
\affiliation{$^1$College of Sciences, North China University of Science and Technology,
Tangshan 063009,  China}
\date{\today}
\begin{abstract}
We analyze the two purely isospin-violating decays $\Lambda_b\rightarrow \Sigma^0 \phi$ and $\Lambda_b\rightarrow \Sigma^0 J/\psi$,
proceeding merely via the exchange topologies, in the framework of the perturbative QCD approach.
Assuming $\Sigma^0$ baryon belongs to the idealized isospin triplet with quark components of $usd$,
the branching ratios of the two decay modes are predicted to be tiny, of the order $10^{-8}-10^{-9}$, leading to difficulty in observing them.
We then extend our study to include the $\Sigma-\Lambda$ mixing.
 It is found that the mixing has significant effect on the $\Lambda_b\rightarrow \Sigma$ decays,
especially it can greatly increase the rate of the $J/\psi$ process, by as much as two orders of magnitude, yield $10^{-7}$,
 which should be searchable in the future.
 We also estimate a set of asymmetry observables with and without the mixing effect, which will be tested in coming experiments.

\end{abstract}



\maketitle

\section{Introduction}
Isospin symmetry is an approximate symmetry in the Standard model (SM) which has been widely used in  the phenomenological analysis of heavy quark decays. 
For example, the amplitudes for the family of $B\rightarrow K\pi$ decays are expected to obey a quadrilateral relation imposed by isospin symmetry 
~\cite{Buras:2003yc,Baek:2004rp,Buras:2003dj,Buras:2004ub},
which allow us to constrain the SM parameters, and to look for signs of new physics (NP).
Based on isospin arguments, the difference between the direct $CP$ asymmetries for the modes $B^+\rightarrow K^+\pi^0$ and $B^0\rightarrow K^+\pi^-$
are expected to be zero at the leading order in the SM~\cite{Gronau:1998ep},
 as the two decays differ only by the spectator quark.
Nevertheless, a nonzero $CP$ asymmetry difference with a significance of more than six standard deviations ($\sigma$)
was observed in the latest measurements of LHCb~\cite{LHCb:2020dpr},
suggests a violation of the strong isospin symmetry may beyond the SM expectation,
which was referred to as the longstanding ``$K\pi$ puzzle"~\cite{Baek:2007yy,Baek:2009pa,Gronau:2005kz,Fleischer:2017vrb,Li:2005kt}.

To resolve above puzzle,
a sizeable electroweak (EW) penguin contribution,  offering an attractive avenue for new particles to enter,
will be requested~\cite{Grossman:1999av,Yoshikawa:2003hb,Mishima:2004um,Gronau:2003kj,Kim:2007kx}.
In the  beauty-hadron decays aspect, the  EW penguin contributions are usually overshadowed by the larger QCD penguins.
For this reason,  it is of particular interest to consider those purely isospin-violating decays,
in which the isospin-conserving QCD penguin amplitudes do not contribute at all.  

The representative examples in hadronic $B$ meson decays
are the $\bar{B}_s\rightarrow (\eta, \eta', \phi) (\pi, \rho)$ modes~\cite{Fleischer:1994rs,Deshpande:1994yd,Chen:1998dta},
whose branching ratios were predicted to be relatively small in the  SM, of order $10^{-8}-10^{-7}$
\cite{Beneke:2003zv,Beneke:2006hg,Ali:2007ff,Zou:2015iwa,Chang:2015wba,Wang:2008rk,Cheng:2009mu,Wang:2017rmh,Faisel:2011kq,Faisel:2013nra,Yan:2017nlj}.
The inclusion of NP effects~\cite{Hofer:2010ee,Faisel:2017glo,Hofer:2011yg,Faisel:2014dna} in EW penguin greatly increases their rates
by an order of magnitude with respect to the SM predictions,
making these decays are interesting for LHCb and future $B$ factories searches.
Up to now,
only the $4\sigma$ evidence for $B^0_s\rightarrow \phi \rho^0$ was seen by LHCb  with a branching fraction of $(2.7\pm0.8)\times10^{-7}$~\cite{LHCb:2016vqn},
which is compatible with its SM expectation albeit still with a significant uncertainty.
This consistency leaves little room for NP contributions
and motivates us to explore further information from other hadronic decays which are also sensitive to EW penguin contributions.

In the $b$-baryon sector, the similar promising processes are the $\Lambda_b\rightarrow \Sigma^0 \phi, \Sigma^0 J/\psi$ decays, which received less attention in the literature.
Since $\Lambda_b$ is an isospin singlet while the final state have total isospin $I=1$, these processes thus also fully break isospin symmetry.
Their amplitudes receive contributions from the tree and EW penguin operators, which are isospin-violating.
By contrast for the QCD penguin operators, which preserve isospin symmetry, do not affect these processes.
In the SM, isospin violation arises from the quark mass difference $m_d-m_u$ and from electromagnetic corrections~\cite{Ecker:2000zr}.
In this case, the QCD-penguin operators can contribute through the isospin-conserving amplitudes if the $\Sigma-\Lambda$
mixing under strong isospin-symmetry breaking is permitted~\cite{Dery:2020lbc}.

The factorization approach (FA) and the diquark approximation failed in describing such purely isospin-violating decays
because the $\Lambda_b\rightarrow \Sigma^0$ transition is forbidden
due to the orthogonality of the  $\Lambda_b$ and $\Sigma^0$ spin functions.
They can only proceed via the exchange diagrams, thus the observations of them would provide valuable information on
the role of exchange topologies in $b$-baryon weak decays.
This poses serious challenges to precise theoretical calculations based on factorization.
However, FA is not necessary in the perturbative QCD (PQCD) approach~\cite{Keum:2000ms}.
Except for the factorizable diagrams,
one can valuate all relevant Feynman diagrams including the exchange topologies in a self-consistent manner.
In PQCD factorization~\cite{Li:1994cka},
the decay amplitude for a exclusive process is obtained as a convolution of a perturbative kernel and  the  nonperturbative hadronic light-cone distribution amplitudes (LCDAs).
After summing the large logarithms associated with parton transverse momenta,
the PQCD formalism with Sudakov suppression can give converging results and have predictive power.
Therefore, it can serve as a useful tool for investigating the heavy bottom decays~\cite{Kurimoto:2001zj,Lu:2002ny,Ali:2007ff,Wang:2010ni,Rui:2021kbn,Chai:2022ptk}. 
Inspired by the success of the PQCD approach in various weak decays of $\Lambda_b$ baryon
\cite{prd59094014,prd61114002,cjp39328,prd65074030,prd74034026,prd80034011,220204804,220209181,prd106053005,221015357,230213785},
in this work we shall perform  a phenomenological analysis of the above two purely isospin-violating decays and explore the exchange topological contribution inside.
The $\Sigma-\Lambda$  mixing effects in the two reactions are also investigated.

The rest of the paper is organized as follows.
We briefly review  the LCDAs, kinematics, and the $\Sigma-\Lambda$ mixing phenomenon in Sec.~\ref{sec:framework}.
The numerical analysis including the results for the decay amplitudes, branching ratios and various asymmetry parameters are presented in Sec.~\ref{sec:results}.
Section~\ref{sec:sum}  is reserved for a summary and conclusions.
Finally, we give the explicit expressions of the factorization formulas in the Appendix.

\section{Theoretical framework}\label{sec:framework}

\subsection{Light-cone distribution amplitudes}\label{sec:LCDAs}
The LCDAs are primary nonperturbative quantities for calculating the heavy baryon decays based on the PQCD approach,
which can be constructed via the nonlocal matrix elements.
We now summarize the definitions of the LCDAs for the initial and final states.

Heavy baryons containing a $b$-quark can be regarded as a system of an effective heavy quark and a pair of light quarks with aligned helicities, called diquark in the heavy quark limit.
The heavy quarks are nonrelativistic particles which decouple from the diquark in the leading order of the heavy quark mass expansion.
The investigation of heavy-baryon distribution amplitudes has made brilliant progress~\cite{plb665197,jhep112013191,epjc732302,plb738334,jhep022016179,Ali:2012zza}
since the complete classification and renormalization group equations (RGE) that govern the scale-dependence are presented~\cite{plb665197}.
Following Ref.~\cite{jhep112013191}, the $\Lambda_b$ baryon LCDAs up to twist-4 accuracy in the momentum space is defined by
\begin{eqnarray}
(\Psi_{\Lambda_b})_{\alpha\beta\gamma}(x_i,\mu)=\frac{1}{8N_c}
\{f_{\Lambda_b}^{(1)}(\mu)[M_1(x_2,x_3)\gamma_5C^T]_{\gamma\beta}
+f_{\Lambda_b}^{(2)}(\mu)[M_2(x_2,x_3)\gamma_5C^T]_{\gamma\beta}\}[\Lambda_b(p)]_\alpha,
\end{eqnarray}
where $\Lambda_b(p)$ is the spinor of the $\Lambda_b$ baryon with on-shell momentum $p$ and $\alpha,\beta,\gamma$ are Dirac indices.
The variables $x_2,x_3$ are the momentum fractions carried by the quarks $u,d$, respectively.
$N_c$ is the number of colors. $C$ is the charge conjugation matrix with the properties $C \gamma_\mu^T C^{-1}=-\gamma_\mu$
and $C \gamma_5^T C^{-1}=\gamma_5$ and the superscript $T$ indicates a transposition in the spinor space.
The color antisymmetrization, which is needed to form a color singlet, is not written out explicitly but implied.
The two couplings $f_{\Lambda_b}^{(1,2)}$ are expected to be equal due to a nonrelativistic constituent quark picture of the $\Lambda_b$~\cite{plb665197}.
For their numerical values we quote the result in~\cite{220204804}:
$f_{\Lambda_b}^{(1)}\approx f_{\Lambda_b}^{(2)}\equiv f_{\Lambda_b}=0.021\pm0.004$ GeV$^3$, at the scale $\mu=1$ GeV.
$M_{1}$ and $M_{2}$ are the chiral-even and -odd projectors, respectively, which read
\begin{eqnarray}
M_1(x_2,x_3)&=&\frac{\slashed{v}\slashed{n}}{4}\Psi_3^{+-}(x_2,x_3)
+\frac{\slashed{n}\slashed{v}}{4}\Psi_3^{-+}(x_2,x_3), \nonumber\\
M_2(x_2,x_3)&=&\frac{\slashed{n}}{\sqrt{2}}\Psi_2(x_2,x_3)
+\frac{\slashed{v}}{\sqrt{2}}\Psi_4(x_2,x_3),
\end{eqnarray}
where $n=(1,0,\textbf{0}_T)$ and $v=(0,1,\textbf{0}_T)$ are dimensionless vectors on the light cone, satisfying $n\cdot v=1$.
The above definition indicates the daughter baryon momentum is along the $n$ direction in the massless limit.
Several asymptotic models for the  various twist LCDAs, in which the twist is indicated by the subscript numbers,
have been proposed in Refs.~\cite{plb665197,jhep112013191,Ali:2012zza}  and summarized in Ref.~\cite{220204804}.
We adopt the exponential model, whose explicit expressions can be found in the previous work~\cite{220204804,prd106053005,221015357} and shall not be repeated here.
The twist 2 and twist 4 LCDAs are symmetric under exchange of $x_2$ and $x_3$,
while the two twist 3 ones do not have any definite symmetry but satisfy  $\Psi_3^{+-}(x_2,x_3)=\Psi_3^{-+}(x_3,x_2)$
because of the isospin zero condition for the diquark.

Similar to the  $\Lambda^0$ case,
the nonlocal matrix element associated with the $\Sigma^0$ baryon is decomposed  into three terms to leading twist accuracy:
\begin{eqnarray}\label{eq:wave}
(\Psi_{\Sigma})_{\alpha\beta\gamma}(k_i',\mu)
&=&\frac{1}{8\sqrt{2}N_c}
\{(\slashed{p}'C)_{\beta\gamma}[\gamma_5\Sigma({p}')]_\alpha\Phi^V(k_i',\mu)
+(\slashed{p}'\gamma_5C)_{\beta\gamma}[\Sigma({p}')]_\alpha\Phi^A(k_i',\mu) \nonumber\\
&&+(i\sigma_{\mu\nu}{p'}^\nu C)_{\beta\gamma}[\gamma^\mu\gamma_5\Sigma({p}')]_\alpha\Phi^T(k_i',\mu)\},
\end{eqnarray}
with $\sigma_{\mu\nu}=i[\gamma_\mu,\gamma_\nu]/2$.
$\Sigma({p}')$ is the $\Sigma$ baryon spinor that satisfies the Dirac equation $\slashed{p}'\Sigma({p}')=m_\Sigma \Sigma({p}')$
with momentum $p'$ and mass $m_\Sigma $.
$\Phi^{V}$, $\Phi^A$ and $\Phi^T$ are the vector, axial-vector and tensor structure  LCDAs, respectively.
Their explicit expressions have been studied using QCD sum rules~\cite{zpc42569,Farrar:1988vz,Liu:2014uha,Liu:2008yg} and lattice QCD~\cite{jhep020702016,prd89094511,epja55116}.
Quite recently, the one-loop perturbative contributions to LCDA of a light baryon has been derived in large-momentum effective theory~\cite{Deng:2023csv},
which provides a first step to obtaining the LCDA from first principle lattice QCD calculations in the future.
In this work, we adopt the Chernyak-Ogloblin-Zhitnitsky (COZ) model for the $\Sigma^0$ baryon LCDAs at the scale $\mu=1$ GeV  proposed in Ref.~\cite{zpc42569}
\begin{eqnarray}\label{eq:vat}
\Phi^V(x'_1,x'_2,x'_3)&=&42f_{\Sigma}\phi_{asy}[0.3({x'_2}^{2}+{x'_3}^2)+0.14{x'_1}^2-0.54x'_2x'_3-0.16x'_1(x'_2+x'_3)],\nonumber\\
\Phi^A(x'_1,x'_2,x'_3)&=&-42f_{\Sigma}\phi_{asy}[0.06({x'_2}^2-{x'_3}^2)+0.05(x'_2-x'_3)],\nonumber\\
\Phi^T(x'_1,x'_2,x'_3)&=&42f_{\Sigma}^T\phi_{asy}[0.32({x'_3}^2+{x'_2}^2)+0.16{x'_1}^2-0.47x'_2x'_3-0.24x'_1(x'_2+x'_3)],
\end{eqnarray}
where  $\phi_{asy}(x'_1,x'_2,x'_3)=120x'_1x'_2x'_3$ denotes the asymptotic form at infinitely large scales.
It can be seen that the LCDAs at the scale $\mu=1$ GeV differ greatly from their asymptotic forms.
It is easy to observe that
$\Phi^{V}$ and $\Phi^{T}$ are symmetric under permutation of $u$ and $d$ quarks,
but $\Phi^A$ is antisymmetric under the same operation.
These symmetry properties are completely opposite to those of the $\Lambda^0$ baryon.
The normalization constant $f_{\Sigma}^{(T)}$ are defined in such a way that~\cite{zpc42569}
\begin{eqnarray}
\int [dx'] \Phi^{V(T)}=f_{\Sigma}^{(T)},
\end{eqnarray}
with the integration measure for the longitudinal momentum fractions:
\begin{eqnarray}
\int [dx'] =\int_0^1  dx'_1dx'_2dx'_3\delta(1-x'_1-x'_2-x'_3).
\end{eqnarray}
The $\delta$ function enforces momentum conservation.
Their values are fixed to be $f_{\Sigma}=0.51\times10^{-2}$ GeV$^2$ and $f^T_{\Sigma}=0.49\times10^{-2}$GeV$^2$~\cite{zpc42569}.

For the $J/\psi$ meson, the longitudinally and transversely polarized LCDAs up to twist-3 are defined by the following expansion~\cite{prd71114008}
\begin{eqnarray}
\Psi_{L}&=&\frac{1}{\sqrt{2N_c}}(m_V\slashed{\epsilon}_L\psi^L+\slashed{\epsilon}_L\slashed{q}\psi^t),\nonumber\\
\Psi_{T}&=&\frac{1}{\sqrt{2N_c}}(m_V\slashed{\epsilon}_T\psi^V+\slashed{\epsilon}_T\slashed{q}\psi^T),
\end{eqnarray}
where $m_V$, $q$ and $\epsilon_{L,T}$ are the mass, momentum, and polarization vectors of vector meson, respectively.
The  expressions of various twists $\psi^{L,T,V,t}$ have been derived~\cite{Sun:2008ew,prd90114030,epjc76564}
 based on the harmonic oscillator wave functions potentials.
\begin{eqnarray}
\psi^{L,T}(y,b)&=&\frac{f_\psi}{2\sqrt{2N_c}}N^{L,T}y\bar{y}
\exp\{-\frac{m_c}{\omega}y\bar{y}[(\frac{y-\bar{y}}{2y\bar{y}})^2+\omega^2b^2]\} ,\nonumber\\
\psi^t(y,b)&=&\frac{f_\psi}{2\sqrt{2N_c}}N^t(y-\bar{y})^2
\exp\{-\frac{m_c}{\omega}y\bar{y}[(\frac{y-\bar{y}}{2y\bar{y}})^2+\omega^2b^2]\},\nonumber\\
\psi^V(y,b)&=&\frac{f_\psi}{2\sqrt{2N_c}}N^V[1+(y-\bar{y})^2]
\exp\{-\frac{m_c}{\omega}y\bar{y}[(\frac{y-\bar{y}}{2y\bar{y}})^2+\omega^2b^2]\},
\end{eqnarray}
with $m_c$ being the charm quark mass. $y$ is the charm quark momentum fraction with the shorthand $\bar y=1-y$,
while $b$ is the corresponding transverse momentum in the $b$ space.
We take the shape parameters $\omega=0.6$ GeV for $J/\psi$~\cite{Sun:2008ew,prd90114030}.
$N^{L,T,V,t}$  are the normalization constants and obey the normalization conditions
\begin{eqnarray}
\int_0^1\psi^{L,T,V,t}(x,0)dx=\frac{f_\psi}{2\sqrt{2N_c}}.
\end{eqnarray}

The longitudinal ($L$) and transverse ($T$) polarizations LCDAs for the vector $\phi$ meson
are written, up to twist 3, as~\cite{Ball:1998sk,Ball:1998ff,Ball:2007rt}
\begin{eqnarray}
\Phi_V^L(y)&=&\frac{1}{\sqrt{2N_c}}[m_V\rlap{/}{\epsilon_L}\phi_V(y)
+\rlap{/}{\epsilon_L}\rlap{/}{q}\phi_V^t(y)+m_V\phi_V^s(y)],\nonumber\\
\Phi_V^T(y)&=&\frac{1}{\sqrt{2N_c}}[m_V\rlap{/}{\epsilon_T}\phi_V^v(y)
+\rlap{/}{\epsilon_T}\rlap{/}{q}\phi_V^T(y)+im_V\epsilon^{\mu\nu\rho\sigma}
\gamma_5\gamma_{\mu}\epsilon_{T\nu}v_{\rho}n_{\sigma}\phi_V^a(y)],
\end{eqnarray}
respectively.
Here, we adopt the convention $\epsilon^{0123}=1$ for the Levi-Civita tensor $\epsilon^{\mu\nu\rho\sigma}$.
The leading-twist distribution amplitude is conventionally expanded in Gegenbauer polynomials,
\begin{eqnarray}\label{eq:twist2}
\phi^{(T)}_V(y)&=&\frac{f^{(T)}_V}{\sqrt{2N_c}}3y(1-y)[1+a_{1V}^{\parallel(\perp)}3(2y-1)+a_{2V}^{\parallel(\perp)}3(5(2y-1)^2-1)/2],\nonumber\\
\end{eqnarray}
where the values of the Gegenbauer moments and decay constants for the $\phi$ meson are taken as~\cite{Ball:2007rt,Rui:2017fje}
\begin{eqnarray}
a_{1\phi}^{\parallel}=a_{1\phi}^{\perp}=0,\quad a_{2\phi}^{\parallel}=0.18\pm 0.08, \quad a_{2\phi}^{\perp}=0.14\pm0.07,\quad f_\phi=(215\pm5)~\text{MeV},\quad f_\phi^T=(186\pm9)~\text{MeV}.
\end{eqnarray}
As for the twist-3 ones, we adopt the asymptotic form:
\begin{eqnarray}
\phi_V^t(y)&=&\frac{3f_V^T}{2\sqrt{2N_c}}(2y-1)^2,\quad \phi_V^s(y)=-\frac{3f_V^T}{2\sqrt{2N_c}}(2y-1),\nonumber\\
\phi_V^v(y)&=&\frac{3f_V}{8\sqrt{2N_c}}[1+(2y-1)^2],\quad \phi_V^a(y)=-\frac{3f_V}{4\sqrt{2N_c}}(2y-1).
\end{eqnarray}

\subsection{ PQCD FORMALISM}
\begin{figure}[!htbh]
	\begin{center}
		\vspace{0.01cm} \centerline{\epsfxsize=15cm \epsffile{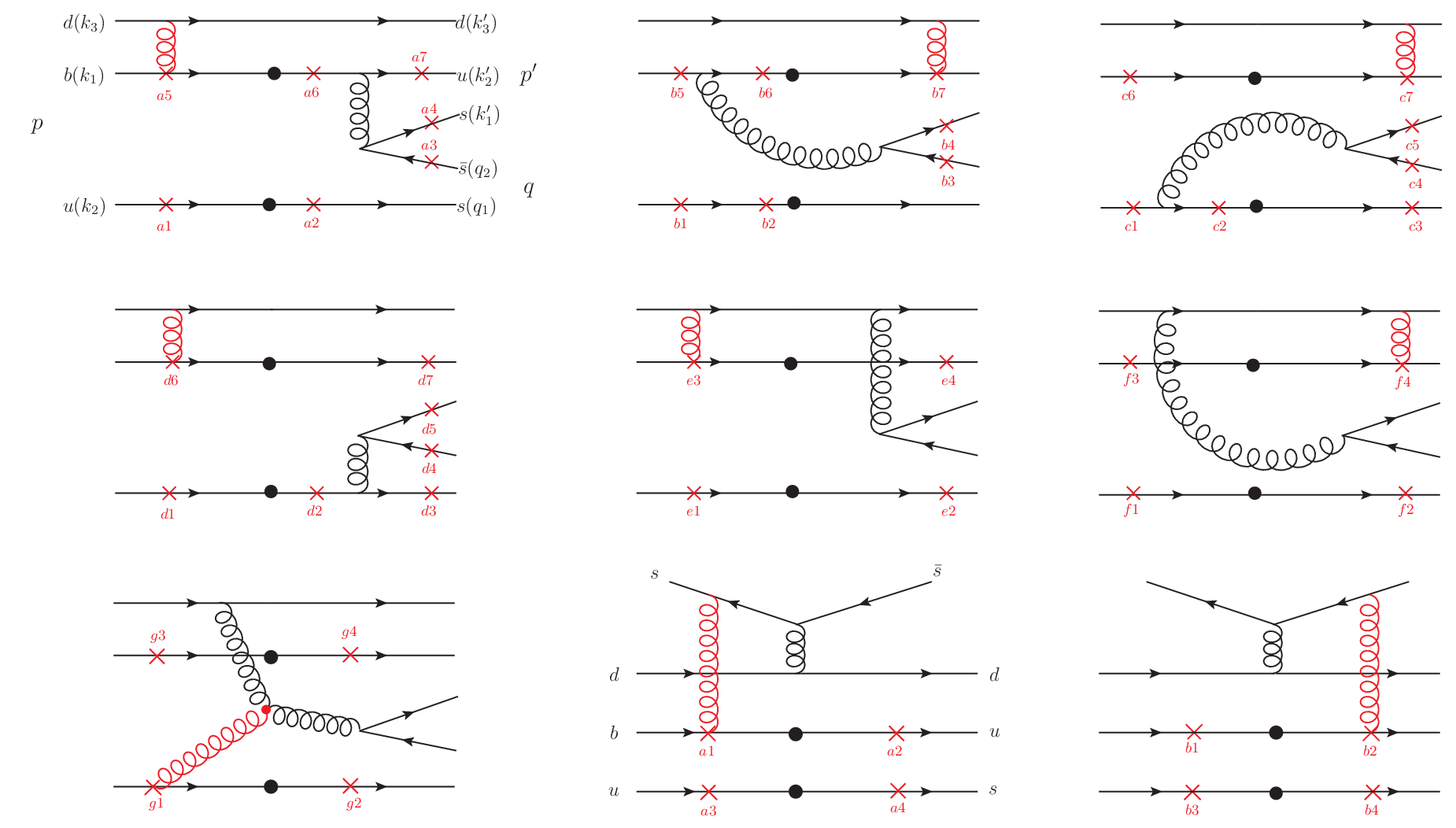}}
		\setlength{\abovecaptionskip}{1.0cm}
		\caption{$W$ exchange diagrams for the decay $\Lambda_b\rightarrow \Sigma^0 \phi$.
The first two rows  are called $E$-type diagrams marked by $E_{ij}$ with $i=a-f$ and $j=1-7$.
The first diagram in the third row is the three-gluon $E$-type marked by $E_{gj}$ with $j=1-4$.
The last two diagrams can be obtained from the W exchange diagram by exchanging the two identical strange quarks in the final states.}
		\label{fig:FeynmanE}
	\end{center}
\end{figure}

As mentioned above, the QCD penguin operators do not give contributions to the purely isospin-violating decays,
so the related effective Hamiltonian can be written as~\cite{Buchalla:1995vs}
\begin{eqnarray}\label{eq:heff}
\mathcal{H}_{eff}&=&\frac{G_F}{\sqrt{2}} \left(\xi_u[C_1(\mu)O^u_1(\mu)+C_2(\mu)O^u_2(\mu)]-\sum_{k=7}^{10} \xi_t C_k(\mu)O_k(\mu)\right)+H.c.,
\end{eqnarray}
where $\xi_{u(t)}=V_{u(t)b}V^*_{u(t)s}$ represent the product of the Cabibbo-Kobayashi-Maskawa (CKM) matrix elements.
$O_{1,2}$ are the so-called current-current operators and $O_{7-10}$ are EW penguin operators, read as
\begin{eqnarray}\label{eq:O7}
O_1&=& \bar{u}_\alpha \gamma_\mu(1-\gamma_5) b_\beta  \otimes \bar{s}_\beta  \gamma^\mu(1-\gamma_5) u_\alpha, \nonumber\\
O_2&=& \bar{u}_\alpha \gamma_\mu(1-\gamma_5) b_\alpha \otimes \bar{s}_\beta  \gamma^\mu(1-\gamma_5) u_\beta,  \nonumber\\
O_7&=&   \frac{3}{2}\bar{s}_\beta \gamma_\mu(1-\gamma_5) b_\beta  \otimes \sum_{q'} e_{q'}\bar{q}'_\alpha \gamma^\mu(1+\gamma_5) q'_\alpha, \nonumber\\
O_8&=&   \frac{3}{2}\bar{s}_\beta \gamma_\mu(1-\gamma_5) b_\alpha \otimes \sum_{q'} e_{q'}\bar{q}'_\alpha \gamma^\mu(1+\gamma_5) q'_\beta,  \nonumber\\
O_9&=&   \frac{3}{2}\bar{s}_\beta \gamma_\mu(1-\gamma_5) b_\beta  \otimes \sum_{q'} e_{q'}\bar{q}'_\alpha \gamma^\mu(1-\gamma_5) q'_\alpha, \nonumber\\
O_{10}&=&\frac{3}{2}\bar{s}_\beta \gamma_\mu(1-\gamma_5) b_\alpha \otimes \sum_{q'} e_{q'}\bar{q}'_\alpha \gamma^\mu(1-\gamma_5) q'_\beta,
\end{eqnarray}
with $\alpha,\beta$ being colors.
The sums over quark flavors in Eq.~(\ref{eq:O7}) run over $u$, $d$, $s$, $c$, and $b$ for the $b$ hadronic decays.
$C_i(\mu)$ are the corresponding Wilson coefficients at the renormalization scale $\mu$, which describe the strength with which a given operator enters the Hamiltonian.

At leading order in $\alpha_s$ expansion,
the $\Lambda_b\rightarrow \Sigma^0\phi$ decay at quark level
occur via the exchange topological diagrams involving two hard gluon exchange in PQCD  demonstrated in Fig.~\ref{fig:FeynmanE}.
The first two rows are classified as $E$-type exchange diagrams marked by $E_{ij}$ with $i=a-f$ and $j=1-7$.
The first diagram in the third row is the $E$-type three-gluon one labeled by $E_{gj}$ with $j=1-4$.
The last two topologies, denoted by $B_{ij}$ with $i=a,b$ and $j=1-4$,
can be obtained from the $E$-type diagrams by exchanging the two identical strange quarks in the final states.
Note that the $s\bar s$ pair must attach two gluons to form a color singlet $\phi$ meson, so only four diagrams contribute to each topology.
As for the $\Lambda_b\rightarrow \Sigma^0 J/\psi$ decay,
the $E$-type diagrams are forbidden
while  $B$-type diagrams can contribute to the decay  by substituting $s\bar s\rightarrow c\bar c$ in Fig~\ref{fig:FeynmanE}.
Now, the decays amplitudes can be derived from these diagrams by inserting the four-quark operators shown in  Eq.~(\ref{eq:O7}).
To match the fermion flows of the diagrams, one must perform the Fierz transformation to the penguin operators,
then both $(V-A)(V-A)$ and $(S-P)(S+P)$ amplitudes contribute to the considered decays.

To derive the decay amplitudes one has to specify the kinematics of initial and final states.
Within the $\Lambda_b$ rest frame, two final states flight back to back.
The $\Lambda_b$ momentum satisfy $p=\frac{M}{\sqrt{2}}(1,1,\textbf{0}_{T})$ with $M$ being the $\Lambda_b$ baryon mass in the light-cone coordinates.
One may further choose the two daughter particle  momenta as
\begin{eqnarray}\label{eq:pq}
p'=\frac{M}{\sqrt{2}}(f^+,f^-,\textbf{0}_{T}), \quad
q=\frac{M}{\sqrt{2}}\left(1-f^+,1-f^-,\textbf{0}_{T}\right),
\end{eqnarray}
such that the $\Sigma^0$ travels in positive and the vector meson in negative direction on the light cone.
These choices are consistent with the definitions of the LCDAs in the previous subsection.
The on-shell conditions $p'^2=m_\Sigma^2$ and $q^2=m_V^2$ for the final-state hadrons lead to
\begin{eqnarray}
f^\pm=\frac{1}{2}\left(1-r_V^2+r_{\Sigma}^2 \pm \sqrt{(1-r_V^2+r_{\Sigma }^2)^2-4r_{\Sigma}^2}\right),
\end{eqnarray}
with the mass ratios $r_{\Sigma,V}=m_{\Sigma,V}/M$.

For the vector meson, the longitudinal and transverse polarization vectors ($\epsilon_{L,T}$) can be determined by
the normalization and orthogonality conditions as
\begin{eqnarray}
\epsilon_L=\frac{1}{\sqrt{2(1-f^+)(1-f^-)}}\left(f^+-1,1-f^-,\textbf{0}_{T}\right),
\quad \epsilon_T=\left(0,0,\textbf{1}_{T}\right).
\end{eqnarray}

For the evaluation of the hard kernels, we need define eight valence quark momenta inside the initial and final states,
which are parametrized as
\begin{eqnarray}
k_1&=&\left(\frac{M}{\sqrt{2}},\frac{M}{\sqrt{2}}x_1,\textbf{k}_{1T}\right),\quad
k_{2,3}=\left(0,\frac{M}{\sqrt{2}}x_{2,3},\textbf{k}_{2T,3T}\right),
k_{1,2,3}'=\left(\frac{M}{\sqrt{2}}f^+x_{1,2,3}',0,\textbf{k}'_{1T,2T,3T}\right),\nonumber\\
q_{1(2)}&=&\left(\frac{M}{\sqrt{2}}y(\bar y)(1-f^+),\frac{M}{\sqrt{2}}y(\bar y)(1-f^-),(-)\textbf{q}_{T}\right).
\end{eqnarray}
Here we keep the parton transverse momenta $\textbf{k}^{(')}_{lT}$  and $\textbf{q}_T$ to smears the endpoint singularities.
Both longitudinal and transverse momenta are subject to the  momentum conservation constraints
\begin{eqnarray}
\sum_{l=1}^3x^{(')}_l=1,\quad \sum_{l=1}^3\textbf{k}^{(')}_{lT}=0.
\end{eqnarray}

By using $\mathcal{H}_{eff}$ in Eq.~(\ref{eq:heff}), the amplitude for the decay of $\Lambda_b$ into a final state $\Sigma^0 V$ is then simply given by
\begin{eqnarray}\label{eq:amp}
\mathcal{M}=\langle \Sigma^0 V|\mathcal{H}_{eff}|\Lambda_b\rangle,
\end{eqnarray}
with $V=\phi,J/\psi$.
As demonstrated in~\cite{prd106053005,221015357},
one can divide the formula Eq.~(\ref{eq:amp}) into two components,
one that contributes only to longitudinally polarized vector meson labeled by $\mathcal{M}^L$,
and the other to transversely polarized one marked by $\mathcal{M}^T$.
The two components can be expanded with the Dirac spinors and polarization vector as
\begin{eqnarray}\label{eq:kq}
\mathcal{M}^L&=&\bar {\Sigma} (p')\epsilon^{\mu*}_{L}[A_1^L\gamma_{\mu}\gamma_5+A_2^L\frac{p'_{\mu}}{M}\gamma_5+B_1^L\gamma_\mu+B_2^L\frac{p'_{\mu}}{M}]\Lambda_b(p),
\nonumber\\
\mathcal{M}^T&=&\bar {\Sigma} (p')\epsilon^{\mu*}_{T}[A_1^T\gamma_{\mu}\gamma_5+B_1^T\gamma_\mu]\Lambda_b(p),
\end{eqnarray}
where $A$ and $B$ respectively stand for the parity-violating and parity-conserving amplitudes.
Their analytic formulas are collected in the Appendix.
Summing over the final state polarizations and averaging over the initial state spins, we can obtain
the total decay amplitude square 
\begin{eqnarray}
|\mathcal{A}|^2=\frac{1}{2}\sum_{\sigma=L,T}|\mathcal{M}^\sigma|^2.
\end{eqnarray}

The helicity amplitudes $H_{\lambda_\Sigma\lambda_V}$ are more convenient to describe the various physical asymmetry observables in the decay angular distribution.
Here, $\lambda_\Sigma=\pm 1/2$ and $\lambda_V=0,\pm1$
are the helicity components of the baryon $\Sigma^0$ and vector meson, respectively.
Angular momentum conservation in the $\Lambda_b$ decay imposes $|\lambda_\Sigma-\lambda_V|\leq\frac{1}{2}$ such that the allowed helicity amplitudes are
$H_{\frac{1}{2}1}$, $H_{-\frac{1}{2}-1}$, $H_{\frac{1}{2}0}$, and $H_{-\frac{1}{2}0}$,
which are related to the invariant amplitudes $A$ and $B$ in Eq.~(\ref{eq:kq}) as~\cite{zpc55659,prd562799}
\begin{eqnarray}\label{eq:helicity}
H_{\pm\frac{1}{2}\pm1}&=&\mp\sqrt{Q_+}A_1^T-\sqrt{Q_-}B_1^T,\nonumber\\
H_{\pm\frac{1}{2}0}&=& \frac{1}{\sqrt{2}m_V}[\pm\sqrt{Q_+}(M-m_\Sigma)A_1^L\mp\sqrt{Q_-}P_cA_2^L+\sqrt{Q_-}(M+m_\Sigma)B_1^L+\sqrt{Q_+}P_cB_2^L],
\end{eqnarray}
where we use the abbreviations $Q_{\pm}=(M\pm m_\Sigma)^2-m_V^2$ and
$P_c=\frac{\sqrt{Q_+Q_-}}{2M}$  is the momentum of the daughter baryon in the rest frame of $\Lambda_b$.
The $H_{\pm\frac{1}{2}\pm1}$ terms in Eq.~(\ref{eq:helicity}) corresponds to the transverse polarizations and the $H_{\pm\frac{1}{2}0}$ ones to the longitudinal ones.
After the helicity amplitudes summation, we also get the total squared amplitude
\begin{eqnarray}\label{eq:hn}
|\mathcal{A}|^2=|H_{\frac{1}{2}1}|^2+|H_{-\frac{1}{2}-1}|^2+|H_{\frac{1}{2}0}|^2+|H_{-\frac{1}{2}0}|^2.
\end{eqnarray}

The two-body decay branching ratio reads
\begin{eqnarray}\label{eq:two}
\mathcal{B}=\frac{P_c\tau_{\Lambda_b}}{8\pi M^2}|\mathcal{A}|^2.
\end{eqnarray}
Some asymmetry observables can be expressed in terms of the helicity amplitudes as (see details in Refs~\cite{plb72427,Gutsche:2013oea,221015357})
\begin{eqnarray}\label{eq:alp}
\alpha_b &=&-|\hat{H}_{\frac{1}{2}1}|^2+|\hat{H}_{-\frac{1}{2}-1}|^2+|\hat{H}_{\frac{1}{2}0}|^2-|\hat{H}_{-\frac{1}{2}0}|^2, \nonumber\\
r_0 &=&|\hat{H}_{\frac{1}{2}0}|^2+|\hat{H}_{-\frac{1}{2}0}|^2, \nonumber\\
r_1 &=&|\hat{H}_{\frac{1}{2}0}|^2-|\hat{H}_{-\frac{1}{2}0}|^2,\nonumber\\
\alpha_{\lambda_\Sigma}&=&|\hat{H}_{\frac{1}{2}0}|^2+|\hat{H}_{\frac{1}{2}1}|^2-|\hat{H}_{-\frac{1}{2}-1}|^2-|\hat{H}_{-\frac{1}{2}0}|^2, \nonumber\\
\alpha_{\lambda_V}&=&|\hat{H}_{\frac{1}{2}0}|^2+|\hat{H}_{-\frac{1}{2}0}|^2-|\hat{H}_{\frac{1}{2}1}|^2-|\hat{H}_{-\frac{1}{2}-1}|^2,
\end{eqnarray}
where the hatted helicity amplitudes $\hat{H}_{\lambda_\Sigma\lambda_V}=H_{\lambda_\Sigma\lambda_V}/|\mathcal{A}|$  are normalized to 1.
The $\alpha_b$  parameter is the parity violating asymmetry characterizing the  decay.
$\alpha_{\lambda_\Sigma}$ represents the longitudinal polarization of the $\Sigma^0$
and $\alpha_{\lambda_V}$ denotes the asymmetry between the longitudinal and transverse polarizations of vector meson.
 $r_0 $ and $r_1$ are the longitudinal unpolarized and polarized parameters, respectively.
Finally,  the direct $CP$ asymmetry of decay  is defined by
\begin{eqnarray}
A_{CP}=\frac{|\mathcal{A}|^2-|\mathcal{\bar{A}}|^2}{|\mathcal{A}|^2+|\mathcal{\bar{A}}|^2},
\end{eqnarray}
where $\mathcal{\bar{A}}$ denotes the decay amplitude of the $CP$ conjugate process.

\subsection{ $\Sigma-\Lambda$  mixing phenomenon}

Isospin symmetry can be broken either by electroweak effects or by the strong interaction through the up and down quark mass difference,
which might lead  to the quark flavor mixing.
In other words,  the observed physical states with definite mass are actually mixtures of the idealized isospin states.
In the SM, $\Lambda^0$ and $\Sigma^0$ baryons are part of the baryon octet and share the same valence quarks $(uds)$.
Their different isospin separates the former as a singlet and the latter as part of the triplet with two partners $\Sigma^{\pm}$.
The physical baryons are made of the mixing of isospin triplet and singlet states
\footnote{We use $\Sigma^0,\Lambda^0$ to stand for the unmixed isospin states and $\Sigma,\Lambda$ to denote the physical particle states.}
\begin{eqnarray}\label{eq:mixing}
\left(
\begin{array}{c}
\Lambda\\
\Sigma
\end{array}
\right)=
\left(
\begin{array}{cc}
\cos \theta & -\sin \theta\\
\sin \theta & \cos \theta
\end{array}
\right)
\left(
\begin{array}{c}
\Lambda^0\\
\Sigma^0
\end{array}
\right),
\end{eqnarray}
with $\theta$ being the mixing angle.
Since the mixing phenomenon in the  $\Lambda-\Sigma$ system was first proposed in the late 1970s,
many fruitful theoretical works on the mixing angle have been made so far
~\cite{Dalitz:1964es,Gal:1967oxu,Isgur:1979ed,Gasser:1982ap,Yagisawa:2001gz,Zhu:1998ai,Radici:2001pq,Aliev:2015ela,Horsley:2014koa,Kordov:2019oer}.
Although the  mixing angle  has not yet been given experimentally,
its value has been estimated to be in the range $(0.55-1.2)^\circ$~\cite{Horsley:2014koa,Aliev:2015ela,Kordov:2019oer,Geng:2020tlx,Franklin:2020bvb}.
The smallness of the mixing angle indicates  the degree of isospin-symmetry breaking is weak.
In general, the initial states of $\Lambda_b$  and $\Sigma_b$, corresponding to  $I=0$ and  $I=1$, are also mixed.
However, the mixing receive an additional suppression from the bottom quark mass~\cite{Franklin:1981rc}.
As a consequence, it is ignored in following analysis.

Considering above mixing effect, the physical decay amplitude for the concerned processes can be written as~\cite{Dery:2020lbc}
\begin{eqnarray}\label{eq:amp1}
\langle \Sigma V |\mathcal{H}_{eff}|\Lambda_b\rangle\approx \theta \langle \Lambda^0 V |\mathcal{H}_{0}|\Lambda_b\rangle+\langle \Sigma^0 V |\mathcal{H}_{1}|\Lambda_b\rangle,
\end{eqnarray}
where $\mathcal{H}_0$ and $\mathcal{H}_1$ refer to isospin conserving and breaking Hamiltonian, respectively.
It can be seen the isospin breaking effect  affect not only the mixing of the states but also the amplitude.
The first term on the right-hand side  (rhs) of Eq.~(\ref{eq:amp1}) conserves isospin and hence receives contributions from both electroweak and QCD-penguin,
but the second term belongs to an isospin-violating transition to which the QCD-penguin component does not participate.
The tree operators, mediated by the $b\rightarrow u$ transition, contribute in both terms but suffer from CKM suppression.
Though the  mixing is small, there is lack of quantitative estimate of its effect.
It is then worthwhile to examine whether the large isospin conserving amplitudes are able to compensate for the tiny mixing,
and to give a sizable impact on the $\Lambda_b\rightarrow\Sigma$ decays.

\section{Numerical results}\label{sec:results}
In the numerical calculations, we use the masses (GeV) as follows~\cite{pdg2022}:
\begin{eqnarray}
M&=&5.6196, \quad  m_{\Sigma}=1.193,  \quad m_b=4.8, \quad  m_\phi=1.019, \quad  m_{J/\psi}=3.097.
\end{eqnarray}
The CKM matrix elements in the Wolfenstein parametrization read
\begin{eqnarray}
(V_{ub},V_{us},V_{tb},V_{ts})=(A\lambda^3(\rho-i\eta),\lambda,1,-A\lambda^2),
\end{eqnarray}
with~\cite{pdg2022}
\begin{eqnarray}
\lambda =0.22650, \quad  A=0.790,  \quad \bar{\rho}=0.141, \quad \bar{\eta}=0.357.
\end{eqnarray}
The lifetime of the $\Lambda_b$ is taken to be 1.464 ps from the particle data group (PDG)~\cite{pdg2022}.
Other nonperturbative parameters appearing in the hadron LCDAs have been specified in the preceding section.

There are three crucial differences between $\Lambda_b\rightarrow \Sigma^0$ and $\Lambda_b\rightarrow \Lambda^0$ decays:
(i)  the former can occur only via the exchange topologies in the SM,
which are power suppressed with respect to the emission ones.
(ii) As the QCD penguin operators do not give contributions to the pure $\Delta I=1$ decays,
the $(S-P)(S+P)$ amplitudes,  stem from the Fierz transformation of the $O_{7,8}$ operators, are strongly suppressed by the corresponding Wilson coefficients.
Our numerical results manifest that their contributions are vanishingly small compared to the $(V-A)(V-A)$ ones.
After taken account of the above two aspects, it is then expected that
the $\Sigma^0$ modes have much small decay amplitudes as shown in Table~\ref{tab:invam}.
(iii) The symmetry properties  of the LCDAs of $\Sigma^0$ under the exchange of the up and down quarks are always opposite to the corresponding terms for $\Lambda^0$,
resulting in different relative contributions from $\Phi^{V}$, $\Phi^A$ and $\Phi^T$.
This distinction will cause some observed quantities to behave differently, as detailed more below.

\begin{table}[!htbh]
	\caption{The invariant amplitudes ($10^{-11}$) for $\Lambda_b\rightarrow \Sigma^0 \phi,\Sigma^0 J/\psi$ decays.
The last column are the total decay amplitude, given in GeV. Only central values are presented here.}
	\label{tab:invam}
	\begin{tabular}[t]{lccccccc}
		\hline\hline
 Mode  & $A_1^L$ & $B_1^L$  & $A_2^L$ & $B_2^L$& $A_1^T$   & $B_1^T$ & $|\mathcal{A}|$    \\ \hline		
 $\Sigma^0 \phi$ &    $-9.6-4.8i$ & $13.6+6.7i$    & $-24.8-0.3i$ & $-29.9-0.7i$ &$-3.3-6.9i$& $5.7+11.3i$ &$2.8\times 10^{-9}$  \\
 $\Sigma^0 J/\psi$ &    $6.8-1.7i$ & $3.9-0.2i$    & $13.9-2.3i$ & $-3.3+2.8i$ &$6.6-1.7i$& $3.7-0.4i$  &$7.4\times 10^{-10}$ \\
		\hline\hline
	\end{tabular}
\end{table}	

In Table~\ref{tab:ampvat},
we compare contributions from three components of LCDAs of $\Sigma^0$ baryon to the invariant amplitudes: $\Phi^V$,$\Phi^A$,and $\Phi^T$.
It is apparent that the $\Phi^A$ component for the concerned decays have the smallest contributions.
This observation differs from that in  $\Lambda_b\rightarrow\Lambda \phi$ decay,
 where a dominant contribution from the $\Phi^A$ component was claimed~\cite{221015357}.
 The reason is that the $\Phi^A$ is antisymmetric under the interchange of the $u$ and $d$ quarks for $\Sigma^0$ but symmetric for $\Lambda^0$.
It is also observed that the relative contributions from $\Phi^V$ and $\Phi^T$ components are different between $\Lambda_b\rightarrow\Sigma^0 \phi$ and $\Lambda_b\rightarrow\Sigma^0 J/\psi$ modes,
which can be understood easily as follows.
The $\Lambda_b\rightarrow\Sigma^0 \phi$ decay receives both the $E$ and $B$-type topological contributions.
The numerical analysis shows that the four $E$-type diagrams $E_{a6}$, $E_{c1}$, $E_{d2}$, and $E_{d4}$ play the
most significant role in the decay amplitude for the $\Lambda_b\rightarrow\Sigma^0 \phi$ transition,
while the contributions from $B$-type exchange diagrams are predicted to be vanishingly small.
As shown in Table~\ref{tab:amp},  for the  $E$-type amplitudes,
the $\Phi^{V}$ and $\Phi^A$ components are down by the  power of $r_\Sigma$ or $r_V$, then the $\Phi^T$ one dominates this process.

\begin{table}
\footnotesize
\caption{Contributions to the decay amplitudes from $\Phi^V $, $\Phi^A $, and $\Phi^T $ components in the $\Sigma^0$ baryon LCDAs.}
\label{tab:ampvat}
\begin{tabular}[t]{lccc}
\hline\hline
Amplitude & $\Phi^V $        & $\Phi^A $        & $\Phi^T $                              \\ \hline
$|\mathcal{A}(\Sigma^0 \phi)|$(GeV) &$1.2\times 10^{-9}$&$3.2\times 10^{-10}$  &$1.9\times 10^{-9}$  \\
$|\mathcal{A}(\Sigma^0 J/\psi)|$(GeV) &$4.4\times 10^{-10}$&$1.7\times 10^{-10}$  &$2.2\times 10^{-10}$  \\
\hline\hline
\end{tabular}
\end{table}

Things are different for the $\Lambda_b\rightarrow\Sigma^0 J/\psi$ decay,
which involves only the $B$-type amplitudes dominated by $B_{a3}$ diagram.
From Table~\ref{tab:ampj},
we note that the leading twist term of $H_{B_{a3}}^{LL}$ is proportional to the factor $r_\Sigma r_V(1-x_3-y)x_1(\Phi^V-\Phi^A)\Psi_2$.
Furthermore, the combination of the two virtual quark propagators in $B_{a3}$ diagram is proportional to
$\frac{1}{(1-x_3-y)x'_3}\frac{1}{x_1x'_3}$  if removing the parton transverse momenta (see Table~\ref{tab:ttt}).
Therefore the amplitude of $B_{a3}$ diagram  behaves like $\frac{1}{x_3'}$\footnote{Note that the LCDAs of $\Sigma^0$ baryon also contain one $x_3'$.} 
and induce enhancement in the endpoint region $x_3'\rightarrow 0$.
Likewise, the $B_{b3}$ diagram also exhibits the similar endpoint behavior.
 Our results indicate its contribution is smaller than the $B_{a3}$ one because an additional subleading term associated with $r_c$
yields a significant destructive correction.
On the contrary, this correction becomes constructive in $H_{B_{a3}}^{LL}$ due to the $r_c$ term flipping sign [see the argument below Eq.~(\ref{eq:bbaa})].
The substantial endpoint enhancement was also observed in the previous PQCD calculations for the  $\Lambda_b\rightarrow p$ transition form factor~\cite{220204804},
which warns us to be much cautious when dropping those power suppression terms especially in the endpoint region.
Numerically, such endpoint enhancement can overcomes the power suppression from $r_\Sigma r_V$,
so that the $B_{a3}$  amplitude dominates the decay.
Viewing Table~\ref{tab:ampj}, it is obvious that  the $\Phi^T$ component is always accompanied by the subleading-twist LCDAs of $\Lambda_b$ baryon,
while the $\Phi^V$ one is associated with the leading-twist one.
The above combined effect can explain why this decay is governed by the $\Phi^V$ component as indicated in Table~\ref{tab:ampvat}.

Although the $B$-type amplitudes are of $\mathcal{O}(r_\Sigma r_V)$, we should note that, for the case of $\Lambda_b\rightarrow\Sigma^0 J/\psi$ decay,
$r_V\sim 0.5$ is not a serious suppression factor.
Moreover, 
 the additional terms connected with the charm quark mass
are  numerically significant in the $\Lambda_b\rightarrow\Sigma^0 J/\psi$ amplitude.
It is therefore not surprising that  $|\mathcal{A}(\Lambda_b\rightarrow\Sigma^0 J/\psi)|$
only fall short by a factor of 4 compared to $|\mathcal{A}(\Lambda_b\rightarrow\Sigma^0 \phi)|$ as shown in the Table~\ref{tab:invam}.
The corresponding branching ratios of the two decay modes are estimated to be small, of the order $10^{-8}-10^{-9}$,
which are difficult to measure experimentally at present.

\begin{table}[!htbh]
	\caption{Helicity amplitudes (GeV) and the magnitude squared of normalized helicity amplitudes  for $\Lambda_b\rightarrow \Sigma^0 \phi,\Sigma^0 J/\psi$ decays.}
	\label{tab:helicity}
	\begin{tabular}[t]{lcccc}
		\hline\hline
		$\lambda_{\Sigma}\lambda_{V}$  &$\frac{1}{2}1$ & $-\frac{1}{2}-1$ & $\frac{1}{2}0$ & $-\frac{1}{2}0$ \\ \hline		
		$H_{\lambda_{\Sigma}\lambda_{\phi}}$ & $(-2.3-i2.1)\times10^{-11}$  & $(-4.7-i9.5)\times10^{-10}$  & $(-9.4+i3.2)\times10^{-10}$   & $(-0.8+i2.3)\times10^{-9}$   \\	
		$|\hat{H}_{\lambda_{\Sigma}\lambda_{\phi}}|^2$  &$1.2\times10^{-4}$  &$0.141$  &$0.123$  &$0.736$\\
$H_{\lambda_{\Sigma}\lambda_{J/\psi}}$ & $(-5.2+i1.2)\times10^{-10}$  & $(2.9-i0.9)\times10^{-10}$
 & $(3.6-i0.3)\times10^{-10}$   & $(-1.3+i1.3)\times10^{-10}$   \\	
       $|\hat{H}_{\lambda_{\Sigma}\lambda_{J/\psi}}|^2$  &$0.529$  &$0.165$  &$0.242$  &$0.064$\\
		\hline\hline
	\end{tabular}
\end{table}	
The numerical results of the helicity amplitudes are displayed in the Table~\ref{tab:helicity}.
As mentioned above the $\Phi^T$ contribution dominates over other two components in the $\phi$ mode.
From Eq.~(\ref{eq:bbaa}), one can see that the $\Phi^T$ term in $A_1^{L,T}$ and $B_1^{L,T}$ has opposite signs but is the same in $A_2^{L}$ and $B_2^{L}$.
This pattern will result in
the constructive (destructive) combination of the four terms in $H_{-\frac{1}{2}0}$ ($H_{\frac{1}{2}0}$) [see Eq.~(\ref{eq:helicity})].
Similarly, we have $H_{-\frac{1}{2}-1}\gg H_{\frac{1}{2}1}$ since the interference between $A_1^{T}$ and $B_1^{T}$
is constructive in $H_{-\frac{1}{2}-1}$ but destructive in $H_{\frac{1}{2}1}$.
Note that the values of $A(B)_2^{L}$  are typically large, so that the longitudinal polarization contribution exceeds the transverse one in $\Lambda_b\rightarrow\Sigma^0 \phi$ decay.
One can observe from Table~\ref{tab:helicity} that the helicity amplitudes of the $\phi$ mode are dominated by $H_{-\frac{1}{2}0}$  which occupies about $73.6\%$ of the full contribution.
The observation is different for the $\Lambda_b\rightarrow\Sigma^0 J/\psi$ decay,
in which the $\Phi^V$ component contributes largest amounts to the invariant amplitudes.
In this case, $A_1^{L,T}$ and $B_1^{L,T}$ have the same signs whereas $A_2^{L}$ and $B_2^{L}$ have opposite signs (see also Eq.~(\ref{eq:bbaa})).
Thus we have $H_{\frac{1}{2}1}\gg H_{-\frac{1}{2}-1}$ and $H_{\frac{1}{2}0}\gg H_{-\frac{1}{2}0}$ as shown in Table~\ref{tab:helicity}.
This pattern indicates transversely polarization contributions will  dominate over the longitudinal one in this channel.

\begin{table}
\footnotesize
\caption{Contributions to the  decay amplitudes from the tree and penguin operators.}
\label{tab:ampTP}
\begin{tabular}[t]{lcc}
\hline\hline
Amplitude & Tree        & Penguin                            \\ \hline
$|\mathcal{A}(\Sigma^0 \phi)|$ (GeV)&$1.57\times 10^{-9}$&$2.1\times 10^{-9}$     \\
$|\mathcal{A}(\Sigma^0 J/\psi)|$ (GeV) &$5.4\times 10^{-10}$&$3.5\times 10^{-10}$  \\
\hline\hline
\end{tabular}
\end{table}

As mentioned earlier, the decay amplitudes at hand receive contributions from tree and EW penguin operators.
For the $b\rightarrow s$ transition,
the product of CKM matrix elements in the tree contributions is suppressed compared to that
in the EW penguin ones, and the suppression factor is $|V_{us}V_{ub}|/|V_{ts}V_{tb}|\sim 0.02$.
On the other hand,
the Wilson coefficient of the dominant EW penguin operator $O_9$ around the $m_b$ scale is also of the order $10^{-2}$~\cite{Lu:2000em}. 
Therefore, the tree and EW penguin operator contributions are in fact similar in magnitude as presented in Table~\ref{tab:ampTP}.
Thus one expects that the comparable tree and EW contributions can enhance the direct $CP$ asymmetry.
From Table~\ref{tab:branching}, one see the $A_{CP}(\Lambda_b\rightarrow \Sigma^0\phi)=5.0\%$ is larger than $A_{CP}(\Lambda_b\rightarrow \Lambda^0\phi)=-1.0\%$~\cite{221015357}.
For the $\Lambda_b\rightarrow \Sigma^0 J/\psi$ mode, as indicated in Table~\ref{tab:wilson} of Appendix~\ref{sec:for},
all contributing Feynman diagrams have the same factor $a^{\sigma}$ containing weak phase information,
implies that the sum of the remnant hard kernels yields an overall strong phase.
In this case the strong phase difference only come from the interference between $(V-A)(V-A)$  and $(S-P)(S+P)$ amplitudes.
As noted previously, the latter is highly suppressed relative to the former and the resultant strong phase difference is rather tiny,
such that  the direct $CP$ asymmetry $A_{CP}(\Lambda_b\rightarrow \Sigma^0 J/\psi)$ is estimated to be less than $\mathcal{O}(1\%)$ in magnitude.
We therefore conclude that any measurement of a sizeable direct $CP$ asymmetry in this decay is an unequivocal signal of NP.

\begin{table}[!htbh]
	\caption{Branching ratios  and various asymmetries for the $\Lambda_b\rightarrow \Sigma^0 V$ decay.
The theoretical errors correspond to the uncertainties due to $\omega=0.40\pm0.04$ GeV and the hard scale $t=(1.0\pm0.2)t$, respectively.}
	\label{tab:branching}
	\begin{tabular}[t]{lccccccc}
		\hline\hline
 Mode  & $\mathcal{B}$   & $A_{CP}(\%)$    & $\alpha_b$ & $\alpha_{\lambda_\Sigma}$ & $\alpha_{\lambda_V}$  &$r_0$  &$r_1$ \\ \hline		
 $\Sigma^0 \phi$ &    $5.8^{+0.3+1.4}_{-1.2-1.0}\times10^{-8}$ & $5.0^{+0.5+7.9}_{-8.6-11.0}$ & $-0.47^{+0.00+0.00}_{-0.10-0.04}$    & $-0.75^{+0.00+0.00}_{-0.15-0.00}$
         & $0.72^{+0.00+0.05}_{-0.05-0.00}$      &$0.86^{+0.00+0.01}_{-0.04-0.00}$ & $-0.61^{+0.00+0.00}_{-0.12-0.03}$ \\
 $\Sigma^0 J/\psi$ &  $2.6^{+0.5+1.1}_{-0.5-0.9}\times10^{-9}$ & $0.8^{+0.0+2.4}_{-2.5-1.9}$& $-0.18^{+0.02+0.01}_{-0.01-0.07}$    & $0.54^{+0.04+0.03}_{-0.00-0.09}$
         & $-0.39^{+0.05+0.00}_{-0.00-0.07}$      &$0.31^{+0.02+0.00}_{-0.00-0.04}$ & $0.18^{+0.02+0.02}_{-0.00-0.08}$ \\
		\hline\hline
	\end{tabular}
\end{table}

\begin{table}[!htbh]
	\caption{The same as Table~\ref{tab:branching} but with the mixing effect including.
The error stems from the mixing angle.}
	\label{tab:branching1}
	\begin{tabular}[t]{lccccccc}
		\hline\hline
 Mode  & $\mathcal{B}$   & $A_{CP}(\%)$    & $\alpha_b$ & $\alpha_{\lambda_\Sigma}$ & $\alpha_{\lambda_V}$  &$r_0$  &$r_1$ \\ \hline		
 $\Sigma\phi$ &    $4.4^{+0.0}_{-0.0}\times10^{-8}$ & $-19.3^{+0.5}_{-0.3}$ & $-0.18^{+0.00}_{-0.01}$    & $-0.64^{+0.00}_{-0.01}$
         & $0.54^{+0.00}_{-0.00}$      &$0.77^{+0.00}_{-0.00}$ & $-0.41^{+0.00}_{-0.01}$ \\
 $\Sigma J/\psi$ &  $3.5^{+0.1}_{-0.1}\times10^{-7}$ & $8.6^{+0.3}_{-0.1}$ & $0.04^{+0.00}_{-0.00}$    & $-0.78^{+0.01}_{-0.00}$
         & $-0.07^{+0.00}_{-0.00}$      &$0.47^{+0.01}_{-0.00}$ & $-0.37^{+0.00}_{-0.01}$ \\
		\hline\hline
	\end{tabular}
\end{table}	
We provide the numerical results for the various asymmetry observables defined by Eq.~(\ref{eq:alp}) in Table~\ref{tab:branching},
where the first and second uncertainties arise
from the shape parameter $\omega_0=0.40\pm0.04$ GeV in the $\Lambda_b$ baryon LCDAs
and hard scale $t$ varying from  $0.8t$  to $1.2t$, respectively.
It is observed that the predicted asymmetries for the two modes seem to be distinctly different.
The reason is, again, the helicity amplitude $H_{-\frac{1}{2}0}$ dominate the $\phi$ mode
while the $H_{\frac{1}{2}1}$ one is more preferred in the  $J/\psi$ channel as explained above.
The pattern of the $\Lambda_b\rightarrow \Sigma^0 \phi$ mode coincides with its $\Lambda^0 $ counterpart,
so the asymmetries between the two modes are close to each other.
However,  the PQCD predictions for the asymmetries in $\Lambda_b\rightarrow \Sigma^0 J/\psi$ are at variance with those in $\Lambda_b\rightarrow \Lambda^0 J/\psi$~\cite{prd106053005}.
Some quantities even possess opposite signs, such as  the longitudinal polarization of the daughter baryon.
Our findings can be compared in future.

We now turn to the results by including the $\Sigma-\Lambda$ mixing effect.
According to Eq.~(\ref{eq:amp1}), the physical (mixed) amplitude are modified into
\begin{eqnarray}
\mathcal{A}(\Lambda_b\rightarrow\Sigma V)=\theta \mathcal{A}(\Lambda_b\rightarrow\Lambda^0 V)+\mathcal{A}(\Lambda_b\rightarrow\Sigma^0 V),
\end{eqnarray}
where the first and second terms correspond to the isospin conserving and violating amplitudes, respectively.
The determination of the mixing angle has been extensive discussed in the literature, whose value still varies in a finite range.
We take the value $\theta=(2.07\pm0.03)\times10^{-2}$ (in radians), which was also considered in~\cite{Geng:2020tlx}.
The central values of the isospin conserving amplitudes of $\Lambda_b\rightarrow\Lambda^0 \phi$
and $\Lambda_b\rightarrow\Lambda^0 J/\psi$ are quoted from our previous work~\cite{prd106053005,221015357}, read
\footnote{These values are not explicitly given in~\cite{prd106053005,221015357}, but one can infer them from the corresponding branching ratios.}
\begin{eqnarray}
|\mathcal{A}(\Lambda_b\rightarrow\Lambda^0 \phi)| &=&3.1\times10^{-8}, \nonumber\\ 
|\mathcal{A}(\Lambda_b\rightarrow\Lambda^0 J/\psi)|&=&4.0\times10^{-7}. 
\end{eqnarray}
As can be seen, the isospin conserving amplitudes are at least an order of magnitude larger than the isospin violating ones and can more or less compensate for the suppression from the mixing angle, implying that the isospin conserving amplitudes may provide potentially non-negligible contributions to the decays of $\Lambda_b$ to the physical final state $\Sigma V$.
Comparing Tables~\ref{tab:branching1} with~\ref{tab:branching} for the $\phi$ mode,
it is clearly seen that the mixing causes distinct differences for the physical observables.
The destructive combination of the two isospin amplitudes is responsible for the decrease in branching ratio,
whereas the interference pattern reverses for the $CP$-conjugated process,
so that the direct $CP$ asymmetry varies from $5.0\%$ without mixing to $-19.3\%$ with mixing.

For the $J/\psi$ process, the isospin conserving amplitude is three orders of magnitude higher than the violating one,
significantly surpassing the suppression from the mixing angle.
As a result  the mixing effect has a strong influence on the $J/\psi$ mode.
It is apparent that the mixing effect can hugely amplify the rate of $\Lambda_b\rightarrow\Sigma J/\psi$ above the estimation without mixing, by as much as two orders of magnitude.
The resulting   asymmetries have also changed drastically.

Experimentally, an upper limit on the isospin amplitude ratio $\mathcal{R}=|A(\Lambda_b\rightarrow \Sigma^0 J/\psi)/A(\Lambda_b\rightarrow \Lambda J/\psi)|$
is measured to be 1/21.8 at $95\%$ confidence level by LHCb~\cite{LHCb:2019aci}.
In the absence of $\Sigma-\Lambda$ mixing, we have the isospin amplitude ratio $\mathcal{R}=0.00185$,
which is far below the experimental upper limit.
Taking into account the mixing, this ratio substantially rises  to 0.0214, reaching half of the upper bound.
This estimate is also in agreement with that in~\cite{Dery:2020lbc} and
supports the assumption made in~\cite{Dery:2020lbc} that the dynamic contribution to $\mathcal{R}$ is much smaller
than the static mixing component within SM.
If an enhancement of such dynamic contribution owing to the NP is excluded,
the observation of a substantial decay rate of $\Lambda_b\rightarrow\Sigma^0 J/\psi$ in the future experiments would be strong evidence of the $\Lambda-\Sigma$ mixing.
It follows that the improved precision measurement of the isospin amplitude ratio is desirable to constrain the mixing angle.
\section{conclusion}\label{sec:sum}
In this work we have studied two purely isospin-violating decays $\Lambda_b\rightarrow \Sigma^0\phi, \Sigma^0J/\psi$,
which are generated via the exchange topologies induced by the $bu \rightarrow su$ transition.
Since $\Lambda_b$ and $\Sigma^0$ belong to different isospin representations,
the $\Lambda_b\rightarrow \Sigma^0$ transition is forbidden in the SM.
The factorization approach and the diquark picture failed in describing such decays.
To estimate the exchange topological contributions, we employ the leading order PQCD formalism, 
which allows us to evaluate all relevant topological diagrams involving two hard gluon exchanges systematically.

We first explore the decays under consideration without the $\Sigma-\Lambda$ mixing.
At the quark level, the decay amplitude for an isospin-violating transition receives contributions from both the tree and electroweak penguins,
whereas the QCD penguin vanishes because it preserves isospin symmetry.
The calculated invariant amplitudes and helicity amplitudes are presented in detail in corresponding Tables.
We compare at some length the distinctive patterns  of the two modes as well as their $\Lambda^0$ counterparts.
It has been demonstrated that the helicity amplitudes $H_{-\frac{1}{2}0}$ and $H_{\frac{1}{2}1}$ respectively dominate the $\phi$ and $J/\psi$ modes,
due to different relative contributions from the $\Sigma^0$ baryon LCDAs.
The obtained branching ratios are typically small, at $10^{-8}-10^{-9}$ level, and the direct $CP$ violations are around several percents.
Furthermore, we predict for the first time a set of asymmetry parameters, which will be tested theoretically and experimentally in the future.

Taking into account the $\Sigma-\Lambda$ mixing effect,
the decays of $\Lambda_b$ to the physical final states involve both the isospin-conserving and isospin-violating contributions.
The PQCD calculations reveal that the former are at least an order of magnitude greater than the latter,
which is sufficient to compensate for the suppression caused by the mixing angle,
so that the influence of mixing is more pronounced for the $\Lambda_b\rightarrow \Sigma$ decays than the $\Lambda_b\rightarrow\Lambda$ ones.
The numerical results manifest that the mixing effect modifies significantly both the decay rates and the asymmetry parameters.
In particular,
it boost the branching ratio of $\Lambda_b\rightarrow \Sigma J/\psi$ to the order of $10^{-7}$, which is encouraging for future measurements.
Therefore, this mode is particularly suitable to investigate the isospin violation and to extract the mixing angle.

\begin{acknowledgments}
We would like to acknowledge Prof. Hsiang-nan Li and  Prof. Yue-Long Shen for helpful discussions.
This work is supported by National Natural Science Foundation
of China under Grants No. 12075086 and  No. 11605060 and the Natural Science Foundation of Hebei Province
under Grants No.A2021209002 and  No.A2019209449.
\end{acknowledgments}

\begin{appendix}
\section{FACTORIZATION FORMULAS}\label{sec:for}

\begin{table} [!htbh]
\footnotesize
	\centering
	\caption{The virtualities of the internal propagators $t_{A,B,C,D}$ for the exchange topological diagrams.}
	\newcommand{\tabincell}[2]{\begin{tabular}{@{}#1@{}}#2\end{tabular}}
	\label{tab:ttt}
	\begin{tabular}[t]{lcccc}
		\hline\hline\
		$R_{ij}$ &  $\frac{t_A}{M^2}$   &$\frac{t_B}{M^2}$  &$\frac{t_C}{M^2}$ &$\frac{t_D}{M^2}$ \\ \hline
		$E_{a1}$ &  $x_3x_3'$   &  $x_1'(y-1)$        &$x_3'(1-x_1)$  &$(x_3'-1)(1-y)$ \\
		$E_{a2}$ &  $x_3x_3'$   &  $x_1'(y-1)$        &$x_3'(x_3-y)$  &$(x_3'-1)(1-y)$  \\
		
		$E_{a3}$ &  $x_3x_3'$   &  $(1-x_2')(x_3+y-1)$        &$x_3'(x_3+y-1)$  &$x_3+y-1$\\
		$E_{a4}$ &  $x_3x_3'$   &  $(1-x_2')(x_3+y-1)$        &$x_3(1-x_2')$  &$x_3+y-1$ \\
		
		$E_{a5}$ &  $x_3x_3'$   &  $x_1'(y-1)$        &$x_3'(1-x_2)+x_2$  &$(x_3'-1)(1-y)$\\
		$E_{a6}$ &  $x_3x_3'$   &  $x_1'(y-1)$        &$x_3+y+1$  &$(x_3'-1)(1-y)$ \\
		
		$E_{a7}$ &  $x_3x_3'$   &  $x_1'(y-1)$        &$x_3(1-x_1')$  &$x_3+y-1$\\
		$E_{b1}$ &  $x_3x_3'$   &  $x_1'(y-1)$        &$x_3'(1-x_1)$  &$x_1'(x_1+y-1)-x_1-y+2$ \\
		
		$E_{b2}$ &  $x_3x_3'$   &  $x_1'(y-1)$        &$x_3'(x_3-y)$  &$x_1'(x_1+y-1)-x_1-y+2$\\
		$E_{b3}$ &  $x_3x_3'$   &  $(1-x_2')(x_3+y-1)$        &$x_3'(x_3+y-1)$  &$x_2'(x_2-y)+1$ \\
		
		$E_{b4}$ &  $x_3x_3'$   &  $(1-x_2')(x_3+y-1)$        &$x_3(1-x_2')$  &$x_2'(x_2-y)+1$ \\
		$E_{b5}$ &  $x_3x_3'$   &  $x_1'(y-1)$        &$x_3'(1-x_2)+x_2$  &$x_2'(x_2-y)+1$ \\
		
		$E_{b6}$ &  $x_3x_3'$   &  $x_1'(y-1)$        &$x_1'(x_1+y-1)-x_1-y+2$  &$x_2'(x_2-y)+1$\\
		$E_{b7}$ &  $x_3x_3'$   &  $x_1'(y-1)$        &$x_3(1-x_1')$  &$x_1'(x_1+y-1)-x_1-y+2$ \\
		
		$E_{c1}$ &  $x_3x_3'$   &  $x_1'(y-1)$        &$x_3'(1-x_1)$  &$(x_1-y)(x_2'-1)$\\
		$E_{c2}$ &  $x_3x_3'$   &  $x_1'(y-1)$        &$x_3'(x_3+y-1)$  &$(x_1-y)(x_2'-1)$ \\
		
		$E_{c3}$ &  $x_3x_3'$   &  $x_1'(y-1)$        &$x_3'(x_3-y)$  &$x_1'(x_2+y-1)$\\
		$E_{c4}$ &  $x_3x_3'$   &  $(1-x_2')(x_3+y-1)$        &$x_3'(x_3+y-1)$  &$(x_1-y)(x_2'-1)$ \\
		
		$E_{c5}$ &  $x_3x_3'$   &  $(1-x_2')(x_3+y-1)$        &$x_3(1-x_2')$  &$(x_1-y)(x_2'-1)$\\
		$E_{c6}$ &  $x_3x_3'$   &  $x_1'(y-1)$        &$x_3'(1-x_2)+x_2$  &$x_1'(x_2+y-1)$ \\
		
		$E_{c7}$ &  $x_3x_3'$   &  $x_1'(y-1)$        &$x_3(1-x_1')$  &$x_1'(x_2+y-1)$\\
		$E_{d1}$ &  $x_3x_3'$   &  $x_1'(y-1)$        &$x_3'(1-x_1)$  &$-x_1'$ \\
		
		$E_{d2}$ &  $x_3x_3'$   &  $x_1'(y-1)$        &$(x_3-1)(1-x_2')$  &$-x_1'$\\
		$E_{d3}$ &  $x_3x_3'$   &  $x_1'(y-1)$        &$x_3'(x_3-y)$  &$(x_3-1)(1-x_2')$ \\
		
		$E_{d4}$ &  $x_3x_3'$   &  $(1-x_2')(x_3+y-1)$        &$x_3'(x_3+y-1)$  &$(x_3-1)(1-x_2')$\\
		$E_{d5}$ &  $x_3x_3'$   &  $(1-x_2')(x_3+y-1)$        &$x_3(1-x_2')$  &$(x_3-1)(1-x_2')$\\
		
		$E_{d6}$ &  $x_3x_3'$   &  $x_1'(y-1)$        &$x_3'(1-x_2)+x_2$  &$-x_1'$\\
		$E_{d7}$ &  $x_3x_3'$   &  $x_1'(y-1)$        &$x_3(1-x_1')$  &$-x_1'$ \\
		
		$E_{e1}$ &  $(1-x_2')(x_3+y-1)$   &  $x_1'(y-1)$        &$(x_1-y)(x_2'-1)$  &$(1-x_2')(y-1)$\\
		$E_{e2}$ &  $(1-x_2')(x_3+y-1)$   &  $x_1'(y-1)$        &$(x_3-1)(1-x_2')$  &$(1-x_2')(y-1)$ \\
		
		$E_{e3}$ &  $(1-x_2')(x_3+y-1)$   &  $x_1'(y-1)$        &$x_2'(x_2-y)+1$  &$(1-x_2')(y-1)$\\
		$E_{e4}$ &  $(1-x_2')(x_3+y-1)$   &  $x_1'(y-1)$        &$x_3+y+1$  &$(1-x_2')(y-1)$ \\
		
		$E_{f1}$ &  $(1-x_2')(x_3+y-1)$   &  $x_1'(y-1)$        &$(x_1-y)(x_2'-1)$  &$x_1'(x_3+y-1)$\\
		$E_{f2}$ &  $(1-x_2')(x_3+y-1)$   &  $x_1'(y-1)$        &$(x_3-1)(1-x_2')$  &$x_1'(x_3+y-1)$ \\
		
		$E_{f3}$ &  $(1-x_2')(x_3+y-1)$   &  $x_1'(y-1)$        &$x_2'(x_2-y)+1$  &$x_1'(x_3+y-1)$\\
		$E_{f4}$ &  $(1-x_2')(x_3+y-1)$   &  $x_1'(y-1)$        &$x_3+y+1$  &$x_1'(x_3+y-1)$ \\
		
		$E_{g1}$ &  $x_1'(y-1)$   &  $(1-x_2')(x_3+y-1)$        &$x_3x_3'$  &$(x_1-y)(x_2'-1)$\\
		$E_{g2}$ &  $x_1'(y-1)$   &$(1-x_2')(x_3+y-1)$          &$x_3x_3'$  &$(x_3-1)(1-x_2')$ \\
		
		$E_{g3}$ &  $x_1'(y-1)$   &  $(1-x_2')(x_3+y-1)$        &$x_3x_3'$  &$x_2'(x_2-y)+1$\\
		$E_{g4}$ &  $x_1'(y-1)$   &  $(1-x_2')(x_3+y-1)$       &$x_3x_3'$  &$x_3+y$ \\		
		
		$B_{a1}$  &  $x_3x_3'$      &  $x_3'(x_3+y-1)$        &$x_3'(x_3-1)$  &$x_2(1-x_3')+1$\\
		$B_{a2}$ &  $x_3x_3'$      &  $x_3'(x_3+y-1)$        &$x_3'(x_3-1)$  &$(x_3-1)(1-x_1')$ \\
		
		$B_{a3}$  &  $x_3x_3'$      &  $x_3'(x_3+y-1)$        &$x_3'(x_3-1)$  &$-x_1x_3'$\\
		$B_{a4}$ &  $x_3x_3'$      &  $x_3'(x_3+y-1)$        &$x_3'(x_3-1)$  &$(x_3-1)(1-x_2')$ \\
		
		$B_{b1}$  &  $x_3x_3'$   &  $x_3'(x_3-y)$        &$x_3'(x_3-1)$  &$x_2(1-x_3')+1$\\
		$B_{b2}$ &  $x_3x_3'$   &  $x_3'(x_3-y)$        &$x_3'(x_3-1)$  &$(x_3-1)(1-x_1')$\\
		
		$B_{b3}$  &  $x_3x_3'$   &  $x_3'(x_3-y)$        &$x_3'(x_3-1)$  &$-x_1x_3'$\\
		$B_{b4}$ &  $x_3x_3'$   &  $x_3'(x_3-y)$        &$x_3'(x_3-1)$  &$(x_3-1)(1-x_2')$ \\
		
		\hline\hline
	\end{tabular}
\end{table}

In PQCD formulism, the factorization formula for the invariant amplitudes in Eq.~(\ref{eq:kq}) 
takes the form~\cite{221015357}
\begin{eqnarray}\label{eq:ab}
A(B)=\frac{f_{\Lambda_b}\pi^2 G_F}{18\sqrt{3}}\sum_{R_{ij}}
\int[\mathcal{D}x][\mathcal{D}b]_{R_{ij}}
\alpha_s^2(t_{R_{ij}})e^{-S_{R_{ij}}}\Omega_{R_{ij}}(b,b',b_q)\sum_{\sigma=LL,SP}a_{R_{ij}}^{\sigma}H^{\sigma}_{R_{ij}}(x,x',y),
\end{eqnarray}
with the integration measure
\begin{eqnarray}
[\mathcal{D}x]=[dx_1dx_2dx_3\delta(1-x_1-x_2-x_3)][dx'_1dx'_2dx'_3\delta(1-x'_1-x'_2-x'_3)]dy,
\end{eqnarray}
where the $\delta$ functions enforce momentum conservation.
The summation extends over all possible diagrams $R_{ij}$.
The hard scale $t_{R_{ij}}$ for each diagram is chosen as the maximal virtuality of internal particles
including the factorization scales in a hard amplitude~\cite{prd80034011}:
\begin{eqnarray}
t_{R_{ij}}=\max(\sqrt{|t_A|},\sqrt{|t_B|},\sqrt{|t_C|},\sqrt{|t_D|},w,w',\frac{1}{b_q}),
\end{eqnarray}
where the factorization scales $w^{(')}$ takes the minimum value of $(\frac{1}{b^{(')}_1},\frac{1}{b^{(')}_2},\frac{1}{b^{(')}_3})$.
The expressions of $t_{A,B,C,D}$ are given in Table~\ref{tab:ttt}.
The Sudakov form factor $S_{R_{ij}}$ can be found in~\cite{prd106053005}.
$\Omega_{R_{ij}}$ is the Fourier transformation of the denominator of the hard amplitude from the $k_T$ space to its conjugate $b$ space.
Their forms and the measure $[\mathcal{D}b]_{R_{ij}}$ can be found in Ref~\cite{221015357}.
$a_{R_{ij}}^{\sigma}$ denotes the product of the CKM matrix elements and the Wilson coefficients,
whose expressions are collected in Table~\ref{tab:wilson}.
The index $\sigma=LL$ denotes 
the $(V-A)(V-A)$ type amplitude by inserting the $O_{1,2,9,10}$ operators in the weak vertex;
and $\sigma=SP$ indicates 
 that an amplitude arises in the Fierz transformation of the $O_{7,8}$ operators.

$H^{\sigma}_{R_{ij}}$ is the numerator of the hard amplitude depending on the spin structure of final state.
In Table~\ref{tab:amp}, we give their expressions in the invariant amplitudes $A_1^{T,L}$ and $A_2^L$ for $\Lambda_b\rightarrow \Sigma^0 \phi$ decay.
The corresponding formulas for those $B$ terms can be obtained by the following replacement:
\begin{eqnarray}\label{eq:bbaa}
B_1^{T,L}=A_1^{T,L}|_{r_\phi\rightarrow -r_\phi,r_\Sigma\rightarrow -r_\Sigma, \Phi^{T}\rightarrow -\Phi^{T}}, \quad
B_2^L=A_2^L|_{r_\phi\rightarrow -r_\phi,r_\Sigma\rightarrow -r_\Sigma,\Phi^{V}\rightarrow -\Phi^{V},\Phi^{A}\rightarrow -\Phi^{A}}.
\end{eqnarray}

The hard functions $H^{\sigma}_{B_{ai}}$ with $i=1,2,3,4$ for $\Lambda_b\rightarrow \Sigma^0 J/\psi$ decay are displayed in Table~\ref{tab:ampj}.
The similar expression for the $H^{\sigma}_{B_{bi}}$  can be obtained from $H^{\sigma}_{B_{ai}}$
by substituting $y\rightarrow 1-y$ and 
changing the sign of the $r_c$ terms.
Because the longitudinal and transverse LCDAs of $J/\psi$ have similar Lorentz  structures,
we can obtain $A_1^T$ from $A_1^L$ by the operation $\psi^L\rightarrow \psi^V,\psi^t\rightarrow \psi^T $.
To obtain the $B$ ones, besides the substitute relations in Eq.~(\ref{eq:bbaa}) still applies, 
one must change the sign of the $r_c$ term.

\begin{table}[H]
\normalsize
\centering
	\caption{The expressions of $a^{LL}$ and $a^{SP}$ in Eq.~(\ref{eq:ab}) for the exchange topological diagrams.} 
	\label{tab:wilson}
	\begin{tabular}[t]{lcc}
		\hline\hline
		$R_{ij}$        &$a^{LL}$           &$a^{SP}$             \\ \hline		
		$E_{a1-a7,e1-e4,f4}$ &$\frac{1}{3}V_{ub}V_{us}^*[C_1-C_2]-\frac{1}{2}V_{tb}V_{ts}^*[C_9-C_{10}]$   &$-\frac{1}{2}V_{tb}V_{ts}^*[C_7-C_8]$\\
								
		$E_{b1,b4,b5,b7}$ &$-\frac{1}{3}V_{ub}V_{us}^*[2C_1+C_2]+\frac{1}{2}V_{tb}V_{ts}^*[2C_9+C_{10}]$      &$\frac{1}{2}V_{tb}V_{ts}^*[2C_7+C_8]$ \\
						
		$E_{b2,b3,b6}$ &$\frac{1}{3}V_{ub}V_{us}^*[\frac{1}{4}C_1-C_2]-\frac{1}{2}V_{tb}V_{ts}^*[\frac{1}{4}C_9-C_{10}]$      &$-\frac{1}{2}V_{tb}V_{ts}^*[\frac{1}{4}C_7-C_8]$ \\
						
		$E_{c1,c5,c6,c7}$ &$\frac{1}{3}V_{ub}V_{us}^*[C_1-2C_2]-\frac{1}{2}V_{tb}V_{ts}^*[C_9-2C_{10}]$      &$-\frac{1}{2}V_{tb}V_{ts}^*[C_7-2C_8]$ \\
						
		$E_{c2,c3,c4}$ &$\frac{1}{3}V_{ub}V_{us}^*[C_1-\frac{1}{4}C_2]-\frac{1}{2}V_{tb}V_{ts}^*[C_9-\frac{1}{4}C_{10}]$      &$-\frac{1}{2}V_{tb}V_{ts}^*[C_7-\frac{1}{4}C_8]$ \\	
					
		$E_{d1,d2,d5,d6,d7}$ &$-\frac{2}{3}V_{ub}V_{us}^*[C_1-C_2]+V_{tb}V_{ts}^*[C_9-C_{10}]$   &$V_{tb}V_{ts}^*[C_7-C_8]$\\	
				
		$E_{d3,d4}$ &$\frac{1}{12}V_{ub}V_{us}^*[C_1-C_2]-\frac{1}{8}V_{tb}V_{ts}^*[C_9-C_{10}]$             &$-\frac{1}{8}V_{tb}V_{ts}^*[C_7-C_8]$ \\
				
		$E_{f1}$ &$\frac{1}{3}V_{ub}V_{us}^*[C_1-\frac{5}{4}C_2]-\frac{1}{2}V_{tb}V_{ts}^*[C_9+\frac{5}{4}C_{10}]$      &$-\frac{1}{2}V_{tb}V_{ts}^*[C_7+\frac{5}{4}C_8]$ \\	
		
		$E_{f2}$ &$-\frac{5}{12}V_{ub}V_{us}^*[C_1-C_2]+\frac{5}{8}V_{tb}V_{ts}^*[C_9-C_{10}]$      &$\frac{5}{8}V_{tb}V_{ts}^*[C_7-C_8]$ \\	
		
		$E_{f3}$ &$\frac{1}{3}V_{ub}V_{us}^*[\frac{5}{4}C_1+C_2]-\frac{1}{2}V_{tb}V_{ts}^*[\frac{5}{4}C_9+C_{10}]$      &$-\frac{1}{2}V_{tb}V_{ts}^*[\frac{5}{4}C_7+C_8]$ \\		
		
		$E_{g1}$ &$-\frac{3}{4}V_{ub}V_{us}^*C_2+\frac{9}{8}V_{tb}V_{ts}^*C_{10}$    & $\frac{9}{8}V_{tb}V_{ts}^*C_8$ \\
		
		$E_{g2}$ &$\frac{3}{4}V_{ub}V_{us}^*[C_1-C_2]-\frac{9}{8}V_{tb}V_{ts}^*[C_9-C_{10}]$      &$-\frac{9}{8}V_{tb}V_{ts}^*[C_7-C_8]$ \\
				
		$E_{g3}$&$\frac{3}{4}V_{ub}V_{us}^*C_1-\frac{9}{8}V_{tb}V_{ts}^*C_9$    & $-\frac{9}{8}V_{tb}V_{ts}^*C_7$ \\
		
        $E_{g4}$ &$0$          &$0$\\

		$B_{a1-a4,b1-b4}$  &$-\frac{1}{4}V_{ub}V_{us}^*[C_1-C_2]+\frac{3}{8}V_{tb}V_{ts}^*[C_9-C_{10}]$   &$\frac{3}{8}V_{tb}V_{ts}^*[C_7-C_8]$\\  	
		\hline\hline
	\end{tabular}
\end{table}

\begin{table}[H]
	\footnotesize
	\centering
	\caption{The expressions of $H^{\sigma}_{R_{ij}}$ in the invariant amplitudes $A_1^{T,L}$ and $A_2^L$ for the $\Lambda_b\rightarrow \Sigma^0\phi$ decay.}
	\newcommand{\tabincell}[2]{\begin{tabular}{@{}#1@{}}#2\end{tabular}}
	\label{tab:amp}
	\begin{tabular}[t]{lccc}
		\hline\hline
		$$  &$\frac{A_1^T}{16M^4}$&$\frac{A_1^L}{16M^4}$&$\frac{A_2^L}{16M^4}$\\ \hline
		$H_{E_{a1}}^{LL}$
		&$\Psi_4x_3'(x_3'-1)r_{\Sigma}(\Phi^A+\Phi^V)\phi_V^T$
		&$\Psi_4x_3'(x_3'-1)r_{\Sigma}(\Phi^A+\Phi^V)\phi_V^t$
		&$2\Psi_4x_3'(x_3'-1)r_{\Sigma}(\Phi^A+\Phi^V)\phi_V^t$\\\\
		
		$H_{E_{a1}}^{SP}$
		&\tabincell{c}{$\Psi_4x_3'(x_3'-1)r_{\Sigma}(\Phi^A-\Phi^V)\phi_V^T-(1-x_1)$\\$(y-1)r_V\Phi^T(\Psi_3 ^{-+}+\Psi_3 ^{+-})(\phi_V^a-\phi_V^v)$}
		&\tabincell{c}{$\Psi_4(x_3'-1)x_3'r_{\Sigma}(\Phi^A-\Phi^V)\phi_V^t+(1-x_1)$\\$(y-1)r_V\Phi^T\phi_V(\Psi_3 ^{-+}+\Psi_3 ^{+-})$}
		&\tabincell{c}{$2(y-1)r_V((1-x_1)\Phi^T(\Psi_3^{-+}+\Psi_3^{+-})$\\$\phi_V-\Psi_4x_3'(\Phi^A+\Phi^V))+2\Psi_4(x_3'-1)$\\$x_3'r_{\Sigma}(\Phi^A-\Phi^V)\phi_V^t$}\\\\
		
		$H_{E_{a2}}^{LL}$
		&\tabincell{c}{$\phi_V^T(\Psi_2(x_3'-1)r_{\Sigma}(y-x_3)(\Phi^A+\Phi^V)$\\$-(y-1)x_3'\Psi_3 ^{+-}(r_{\Sigma}-1)\Phi^T)$}
		&\tabincell{c}{$\phi_V^t(\Psi_2(x_3'-1)r_{\Sigma}(y-x_3)(\Phi^A+\Phi^V)$\\$-(y-1)x_3'\Psi_3 ^{+-}(r_{\Sigma}-1)\Phi^T)$}
		&\tabincell{c}{$2\phi_V^t(\Psi_2(x_3'-1)r_{\Sigma}(y-x_3)(\Phi^A+\Phi^V)$\\$+(y-1)x_3'\Psi_3 ^{+-}\Phi^T)$}\\\\
		
		$H_{E_{a2}}^{SP}$
		&\tabincell{c}{$(y-1)\Psi_3^{+-}r_V\Phi^T(y-x_3)(\phi_V^a-\phi_V^v)$\\$+(x_3'-1)r_{\Sigma}(\Phi^A-\Phi^V)$\\$\phi_V^T(\Psi_4x_3'+\Psi_2(y-x_3))$}
		&\tabincell{c}{$(x_3'-1)r_{\Sigma}(\Phi^A-\Phi^V)\phi_V^t$\\$(\Psi_4x_3'+\Psi_2(y-x_3))+(y-1)$\\$\Psi_3 ^{+-}r_V\Phi^T\phi_V(x_3-y)$}
		&\tabincell{c}{$2((x_3'-1)r_{\Sigma}(\Phi^A-\Phi^V)\phi_V^t(\Psi_4x_3'+\Psi_2$\\$(y-x_3))+(y-1)r_V\phi_V((x_3-y)(\Psi_2(\Phi^A$\\$+\Phi^V)+\Psi_3^{+-}\Phi^T)-\Psi_4x_3'(\Phi^A+\Phi^V)))$}\\\\

		$H_{E_{a3}}^{LL}$
		&$\Psi_2r_{\Sigma}(x_3+y-1)(\Phi^A+\Phi^V)\phi_V^T$
		&$\Psi_2r_{\Sigma}(x_3+y-1)(\Phi^A+\Phi^V)\phi_V^t$
		&$2\Psi_2r_{\Sigma}(x_3+y-1)(\Phi^A+\Phi^V)\phi_V^t$\\\\		
		
		$H_{E_{a3}}^{SP}$
		&\tabincell{c}{$(x_3+y-1)(\Psi_3^{-+}r_V\Phi^T(x_3+y-1)$\\$(\phi_V^a-\phi_V^v)+\Psi_2r_{\Sigma}(\Phi^A-\Phi^V)\phi_V^T)$}
		&\tabincell{c}{$(x_3+y-1)(\Psi_2r_{\Sigma}(\Phi^A-\Phi^V)\phi_V^t$\\$-\Psi_3 ^{-+}r_V\Phi^T\phi_V(x_3+y-1))$}
		&\tabincell{c}{$2(x_3+y-1)(r_V\phi_V(x_3+y-1)$\\$(\Psi_2(\Phi^A+\Phi^V)-\Psi_3 ^{-+}\Phi^T)+\Psi_2r_{\Sigma}$\\$(\Phi^A-\Phi^V)\phi_V^t)$}\\\\
		
		$H_{E_{a4}}^{LL}$
		&\tabincell{c}{$x_3\phi_V^T(\Psi_2r_{\Sigma}(\Phi^A+\Phi^V)$\\$-\Psi_3 ^{+-}(r_{\Sigma}-1)\Phi^T)$}
		&\tabincell{c}{$x_3\phi_V^t(\Psi_2r_{\Sigma}(\Phi^A+\Phi^V)$\\$-\Psi_3 ^{+-}(r_{\Sigma}-1)\Phi^T)$}
		&\tabincell{c}{$-2x_3(\Psi_4r_V\phi_V(\Phi^A-\Phi^V)-\phi_V^t$\\$(\Psi_2r_{\Sigma}(\Phi^A+\Phi^V)+\Psi_3 ^{+-}\Phi^T))$}\\\\		
		
		$H_{E_{a4}}^{SP}$
		&$x_3\Psi_2r_{\Sigma}(\Phi^A-\Phi^V)\phi_V^T$
		&$x_3\Psi_2r_{\Sigma}(\Phi^A-\Phi^V)\phi_V^t$
		&\tabincell{c}{$2x_3(\Psi_2r_{\Sigma}(\Phi^A-\Phi^V)\phi_V^t$\\$+\Psi_3 ^{+-}r_V\Phi^T\phi_V)$}\\\\
		
		$H_{E_{a5}}^{LL}$
		&\tabincell{c}{$(x_3'-1)r_{\Sigma}(-(\Phi^A+\Phi^V))$\\$(\Psi_4x_3'+(x_2-1)\Psi_2)\phi_V^T$}
		&\tabincell{c}{$(x_3'-1)r_{\Sigma}(-(\Phi^A+\Phi^V))$\\$(\Psi_4x_3'+(x_2-1)\Psi_2)\phi_V^t$}
		&\tabincell{c}{$-2(x_3'-1)r_{\Sigma}(\Phi^A+\Phi^V)$\\$(\Psi_4x_3'+(x_2-1)\Psi_2)\phi_V^t$}\\\\
		
		$H_{E_{a5}}^{SP}$
		&\tabincell{c}{$-(y-1)\Psi_3^{-+}r_V\Phi^T(\phi_V^a-\phi_V^v)-\Psi_2$\\$(x_2-1)(x_3'-1)r_{\Sigma}(\Phi^A-\Phi^V)\phi_V^T$}
		&\tabincell{c}{$(y-1)\Psi_3^{-+}r_V\Phi^T\phi_V-(x_2-1)\Psi_2$\\$(x_3'-1)r_{\Sigma}(\Phi^A-\Phi^V)\phi_V^t$}
		&\tabincell{c}{$-2((y-1)r_V\phi_V(\Psi_2(\Phi^A+\Phi^V)+\Phi^T$\\$(x_3'\Psi_3^{+-}-\Psi_3^{-+}))+(x_2-1)\Psi_2$\\$(x_3'-1)r_{\Sigma}(\Phi^A-\Phi^V)\phi_V^t)$}\\\\
		
		$H_{E_{a6}}^{LL}$
		&$(y-1)\Psi_3^{+-}(r_{\Sigma}-1)\Phi^T\phi_V^T$
		&$(y-1)\Psi_3^{+-}(r_{\Sigma}-1)\Phi^T\phi_V^t$
		&$-2(y-1)\Psi_3^{+-}\Phi^T\phi_V^t$\\\\
		
		$H_{E_{a6}}^{SP}$
		&\tabincell{c}{$(y-1)\Psi_3^{-+}r_V\Phi^T(x_3+y-1)(\phi_V^a-\phi_V^v)$}
		&$-(y-1)\Psi_3^{-+}r_V\Phi^T\phi_V(x_3+y-1)$
		&\tabincell{c}{$2(y-1)r_V\phi_V(\Psi_2(x_3+y-1)(\Phi^A+\Phi^V)$\\$-\Phi^T(\Psi_3^{-+}(x_3+y-1)+\Psi_3^{+-}))$}\\\\
		
		$H_{E_{a7}}^{LL}$
		&$0$
		&$0$
		&$2 x_3 \Psi _4 r_V \phi _V (\Phi ^A-\Phi^V)$\\\\		
		
		$H_{E_{a7}}^{SP}$
		&$0$
		&$0$
		&$0$\\\\
		
		$H_{E_{b1}}^{LL}$
		&\tabincell{c}{$(y-x_2-x_3)(\Phi^A-\Phi^V)(\phi_V^a+\phi_V^v)$\\$\Psi_4x_3'r_V+\phi_V^T((x_1-1)x_1'(r_{\Sigma}-1)\Phi^T$\\$(\Psi_3^{-+}+\Psi_3^{+-})-\Psi_4x_3'r_{\Sigma}(\Phi^A+\Phi^V))$}
		&\tabincell{c}{$((x_1-1)x_1'(r_{\Sigma}-1)\Phi^T(\Psi_3^{-+}+\Psi_3^{+-})$\\$\phi_V^t-\Psi_4x_3'r_{\Sigma}(\Phi^A+\Phi^V))+\Psi_4x_3'r_V$\\$\phi_V(y-x_2-x_3)(\Phi^A-\Phi^V)$}
		&\tabincell{c}{$2\phi_V^t(\Psi_4x_3'r_{\Sigma}(-(\Phi^A+\Phi^V))$\\$-(x_1-1)x_1'\Phi^T(\Psi_3 ^{-+}+\Psi_3 ^{+-}))$\\$+2\Psi_4x_3'r_V\phi_V(\Phi^A-\Phi^V)$}\\\\	
		
		$H_{E_{b1}}^{SP}$
		&\tabincell{c}{$(\Psi_4x_3'(y+x_1-1)(\Phi^A+\Phi^V)(\phi_V^a+\phi_V^v)$\\$r_V+x_3\Phi^T(2-x_1-y)(\Psi_3^{-+}+\Psi_3^{+-})$\\$(\phi_V^a-\phi_V^v)+x_2\Phi^T(2-x_1-y)$\\$(\Psi_3^{-+}+\Psi_3^{+-})(\phi_V^a-\phi_V^v))$\\$-\Psi_4x_3'r_{\Sigma}(\Phi^A-\Phi^V)\phi_V^T$}
		&\tabincell{c}{$r_V\phi_V(\Psi_4x_3'(y-x_2-x_3)(\Phi^A+\Phi^V)$\\$-(1-x_1)\Phi^T(2-x_1-y)(\Psi_3^{-+}+\Psi_3^{+-}))$\\$-\Psi_4x_3'r_{\Sigma}(\Phi^A-\Phi^V)\phi_V^t$}
		&\tabincell{c}{$2r_V\phi_V(\Psi_4x_3'(\Phi^A+\Phi^V)-(1-x_1)$\\$\Phi^T(2-x_1-y)(\Psi_3 ^{-+}+\Psi_3 ^{+-}))$\\$-2\Psi_4x_3'r_{\Sigma}(\Phi^A-\Phi^V)\phi_V^t$}\\\\
		
		$H_{E_{b2}}^{LL}$
		&\tabincell{c}{$(y-x_3)(\Psi_2r_V(x_1+y-1)(\Phi^A-\Phi^V)$\\$(\phi_V^a+\phi_V^v)+\phi_V^T(\Psi_3^{-+}x_1'(r_{\Sigma}-1)(-\Phi^T)$\\$-\Psi_2r_{\Sigma}(\Phi^A+\Phi^V)))	$}
		&\tabincell{c}{$(y-x_3)(\phi_V^t(\Psi_3^{-+}x_1'(r_{\Sigma}-1)(-\Phi^T)$\\$-\Psi_2r_{\Sigma}(\Phi^A+\Phi^V))+\Psi_2r_V$\\$\phi_V(x_1+y-1)(\Phi^A-\Phi^V))$}
		&\tabincell{c}{$2(y-x_3)(\phi_V^t(\Psi_3^{-+}x_1'\Phi^T-\Psi_2r_{\Sigma}$\\$(\Phi^A+\Phi^V))+\Psi_2r_V\phi_V(\Phi^A-\Phi^V))$}\\\\
		
		$H_{E_{b2}}^{SP}$
		&\tabincell{c}{$(\Psi_2(x_3-y)-\Psi_4x_3')(r_{\Sigma}(\Phi^A-\Phi^V)\phi_V^T$\\$-r_V(x_1+y-1)(\Phi^A+\Phi^V)(\phi_V^a+\phi_V^v))$\\$+\Psi_3 ^{+-}r_V\Phi^T(x_1+y-2)(y-x_3)(\phi_V^a-\phi_V^v)$}
		&\tabincell{c}{$(\Psi_4x_3'+\Psi_2(y-x_3))(r_{\Sigma}(\Phi^V-\Phi^A)$\\$\phi_V^t+r_V\phi_V(x_1+y-1)(\Phi^A+\Phi^V))$\\$+\Psi_3^{+-}r_V\Phi^T\phi_V(x_1+y-2)(x_3-y)$}
		&\tabincell{c}{$2r_V\phi_V((y-x_3)(\Psi_2(\Phi^A+\Phi^V)-\Psi_3^{+-}$\\$\Phi^T(x_1+y-2))+\Psi_4x_3'(\Phi^A+\Phi^V))$\\$-2r_{\Sigma}(\Phi^A-\Phi^V)\phi_V^t(\Psi_4x_3'+\Psi_2(y-x_3))$}\\
		\hline\hline
	\end{tabular}
\end{table}

\begin{table}[H]
	\footnotesize
	\centering
	TABLE~\ref{tab:amp} (continued)
	\newcommand{\tabincell}[2]{\begin{tabular}{@{}#1@{}}#2\end{tabular}}
		\begin{tabular}[t]{lccc}
		\hline\hline
		$$  &$\frac{A_1^T}{16M^4}$&$\frac{A_1^L}{16M^4}$&$\frac{A_2^L}{16M^4}$\\ \hline
		$H_{E_{b3}}^{LL}$
		&\tabincell{c}{$\phi_V^T(r_{\Sigma}(\Psi_2(x_3+y-1)(\Phi^A+\Phi^V)$\\$+\Phi^T(x_3'\Psi_3^{+-}(y-x_2)-\Psi_3^{-+}(x_2'-1)$\\$(x_3+y-1)))+\Phi^T(\Psi_3^{-+}(x_2'-1)$\\$(x_3+y-1)+x_3'\Psi_3 ^{+-}(x_2-y)))$}
		&\tabincell{c}{$\phi_V^t(r_{\Sigma}(\Psi_2(x_3+y-1)(\Phi^A+\Phi^V)$\\$+\Phi^T(x_3'\Psi_3^{+-}(y-x_2)-\Psi_3^{-+}(x_2'-1)$\\$(x_3+y-1)))+\Phi^T(\Psi_3^{-+}(x_2'-1)$\\$(x_3+y-1)+x_3'\Psi_3 ^{+-}(x_2-y)))$}
		&\tabincell{c}{$2(\phi_V^t(\Psi_2r_{\Sigma}(x_3+y-1)(\Phi^A+\Phi^V)$\\$+\Phi^T(\Psi_3^{-+}(x_2'-1)(x_3+y-1)$\\$+x_3'\Psi_3^{+-}(x_2-y)))+r_V\phi_V(\Phi^A-\Phi^V)$\\$(\Psi_4x_3'+\Psi_2(y-x_2-1)(x_3+y-1)))$}\\\\	
		
		$H_{E_{b3}}^{SP}$
		&\tabincell{c}{$(x_3+y-1)(\Psi_3^{-+}r_V\Phi^T(y-x_2-1)$\\$(\phi_V^a-\phi_V^v)+\Psi_2r_{\Sigma}(\Phi^A-\Phi^V)\phi_V^T)$}
		&\tabincell{c}{$(1-x_3-y)(\Psi_3^{-+}r_V\Phi^T\phi_V$\\$(y-x_2-1)-\Psi_2r_{\Sigma}(\Phi^A-\Phi^V)\phi_V^t)$}
		&\tabincell{c}{$2(((y-x_2-1)(x_3+y-1)(\Psi_2(\Phi^A+\Phi^V)$\\$-\Psi_3^{-+}\Phi^T)r_V\phi_V-x_3'\Psi_3^{+-}\Phi^T(y-x_2))$\\$+\Psi_2r_{\Sigma}(x_3+y-1)(\Phi^A-\Phi^V)\phi_V^t)$}\\\\
	
		$H_{E_{b4}}^{LL}$
		&\tabincell{c}{$x_3\Psi_4(x_2'-1)r_V(\Phi^A-\Phi^V)(\phi_V^a+\phi_V^v)$\\$+r_{\Sigma}(\Phi^A+\Phi^V)(x_3\Psi_2-\Psi_4(x_2'-1)^2)$\\$\phi_V^T-x_3\Psi_3 ^{+-}(r_{\Sigma}-1)\Phi^T\phi_V^T$}
		&\tabincell{c}{$r_{\Sigma}(\Phi^A+\Phi^V)(x_3\Psi_2-\Psi_4(x_2'-1)^2)$\\$\phi_V^t+x_3\Psi_4(x_2'-1)r_V\phi_V(\Phi^A-\Phi^V)$\\$-x_3\Psi_3 ^{+-}(r_{\Sigma}-1)\Phi^T\phi_V^t$}
		&\tabincell{c}{$2\phi_V^t(r_{\Sigma}(\Phi^A+\Phi^V)(x_3\Psi_2-\Psi_4(x_2'-1)^2)$\\$+x_3\Psi_3 ^{+-}\Phi^T)-2x_3\Psi_4r_V\phi_V(\Phi^A-\Phi^V)$}\\\\	
		
		$H_{E_{b4}}^{SP}$
		&$x_3\Psi_2r_{\Sigma}(\Phi^A-\Phi^V)\phi_V^T$
		&$x_3\Psi_2r_{\Sigma}(\Phi^A-\Phi^V)\phi_V^t$
		&$2x_3(\Psi_2r_{\Sigma}(\Phi^A-\Phi^V)\phi_V^t+\Psi_3 ^{+-}r_V\Phi^T\phi_V)$\\\\
		
		$H_{E_{b5}}^{LL}$
		&\tabincell{c}{$(\Phi^A-\Phi^V)(x_2\Psi_4+(x_2-1)\Psi_2(x_2-y))$\\$r_V(\phi_V^a+\phi_V^v)+r_{\Sigma}(-(\Phi^A+\Phi^V))$\\$(\Psi_2-\Psi_4x_2'x_3')\phi_V^T-\Psi_3 ^{-+}$\\$((x_2-1)x_2'+1)(r_{\Sigma}-1)\Phi^T\phi_V^T$}
		&\tabincell{c}{$(-(\Phi^A+\Phi^V))(\Psi_2-\Psi_4x_2'x_3')$\\$r_{\Sigma}\phi_V^t+r_V\phi_V(\Phi^A-\Phi^V)$\\$(x_2\Psi_4+(x_2-1)\Psi_2(x_2-y))-\Psi_3 ^{-+}$\\$((x_2-1)x_2'+1)(r_{\Sigma}-1)\Phi^T\phi_V^t$}
		&\tabincell{c}{$2(\phi_V^t(\Psi_3 ^{-+}((x_2-1)x_2'+1)\Phi^T-r_{\Sigma}$\\$(\Phi^A+\Phi^V)(\Psi_2-\Psi_4x_2'x_3'))+(\Phi^A-\Phi^V)$\\$r_V\phi_V(x_2\Psi_4+\Psi_2(x_2^2-x_2(y+1)+1)))$}\\\\	
		
		$H_{E_{b5}}^{SP}$
		&\tabincell{c}{$r_V((y-1)\Psi_3 ^{-+}\Phi^T(\phi_V^v-\phi_V^a)$\\$-(x_2-1)\Psi_2(y-x_2)(\Phi^A+\Phi^V)$\\$(\phi_V^a+\phi_V^v))-\Psi_2r_{\Sigma}(\Phi^A-\Phi^V)\phi_V^T$}
		&\tabincell{c}{$((x_2-1)\Psi_2(x_2-y)(\Phi^A+\Phi^V)$\\$+(y-1)\Psi_3 ^{-+}\Phi^T)r_V\phi_V-\Psi_2r_{\Sigma}$\\$(\Phi^A-\Phi^V)\phi_V^t$}
		&\tabincell{c}{$2r_V\phi_V(\Psi_2(-x_2(y+1)+x_2^2+1)(\Phi^A+\Phi^V)$\\$+\Phi^T((y-1)\Psi_3 ^{-+}+\Psi_3 ^{+-}(x_3'(x_2-y)-x_2)))$\\$-2\Psi_2r_{\Sigma}(\Phi^A-\Phi^V)\phi_V^t$}\\\\
		
		$H_{E_{b6}}^{LL}$
		&\tabincell{c}{$r_V(\Phi^A-\Phi^V)(\phi_V^a+\phi_V^v)((x_1+y-1)$\\$(\Psi_4x_2'+\Psi_2(y-x_2))-\Psi_4)+r_{\Sigma}$\\$(-(\Phi^A+\Phi^V))\phi_V^T(\Psi_2(y-x_2)-\Psi_4x_3')$\\$-(r_{\Sigma}-1)\Phi^T\phi_V^T(\Psi_3 ^{-+}(x_1+y-2)$\\$+\Psi_3 ^{+-}((1-x_1')(y-x_2)-x_3-1))$}
		&\tabincell{c}{$-r_{\Sigma}(\Phi^A+\Phi^V)\phi_V^t(\Psi_2(y-x_2)$\\$-\Psi_4x_3')+r_V\phi_V(\Phi^A-\Phi^V)((x_1+y-1)$\\$(\Psi_4x_2'+\Psi_2(y-x_2))-\Psi_4)-(r_{\Sigma}-1)$\\$\Phi^T\phi_V^t(\Psi_3 ^{-+}(x_1+y-2)$\\$+\Psi_3 ^{+-}((1-x_1')(y-x_2)-x_3-1))$}
		&\tabincell{c}{$2(r_{\Sigma}(-(\Phi^A+\Phi^V))(\Psi_2(y-x_2)$\\$-\Psi_4x_3')(\phi_V^t+r_V\phi_V(\Phi^A-\Phi^V))$\\$+\Phi^T\phi_V^t(\Psi_3 ^{-+}(x_1+y-2)+\Psi_3 ^{+-}$\\$((1-x_1')(y-x_2)-x_3-1)))$}\\\\	
		
		$H_{E_{b6}}^{SP}$
		&\tabincell{c}{$(y-x_2)(r_V(\Psi_2(x_1+y-1)(\Phi^A+\Phi^V)$\\$(\phi_V^a+\phi_V^v)+\Psi_3 ^{-+}\Phi^T(x_1+y-2)$\\$(\phi_V^v-\phi_V^a))-\Psi_2r_{\Sigma}(\Phi^A-\Phi^V)\phi_V^T)$}
		&\tabincell{c}{$(y-x_2)(r_V\phi_V(\Psi_2(x_1+y-1)$\\$(\Phi^A+\Phi^V)+\Psi_3 ^{-+}\Phi^T(x_1+y-2))$\\$-\Psi_2r_{\Sigma}(\Phi^A-\Phi^V)\phi_V^t)$}
		&\tabincell{c}{$2(y-x_2)(r_V\phi_V(\Psi_2(\Phi^A+\Phi^V)$\\$+\Psi_3 ^{-+}\Phi^T(x_1+y-2))$\\$-\Psi_2r_{\Sigma}(\Phi^A-\Phi^V)\phi_V^t)$}\\\\
		
		$H_{E_{b7}}^{LL}$
		&$0$
		&$0$
		&$0$\\\\	
		
		$H_{E_{b7}}^{SP}$
		&$0$
		&$0$
		&$0$\\\\
		
		$H_{E_{c1}}^{LL}$
		&\tabincell{c}{$(x_2'-1)\phi_V^T(\Psi_4x_3'r_{\Sigma}(-(\Phi^A+\Phi^V))$\\$-(1-x_1)(r_{\Sigma}-1)\Phi^T(\Psi_3^{-+}+\Psi_3 ^{+-}))$}
		&\tabincell{c}{$(x_2'-1)\phi_V^t(\Psi_4x_3'r_{\Sigma}(-(\Phi^A+\Phi^V))$\\$-(1-x_1)(r_{\Sigma}-1)\Phi^T(\Psi_3^{-+}+\Psi_3^{+-}))$}
		&\tabincell{c}{$2(x_2'-1)\phi_V^t((1-x_1)\Phi^T(\Psi_3^{-+}+\Psi_3^{+-})$\\$-\Psi_4x_3'r_{\Sigma}(\Phi^A+\Phi^V))$\\$+2\Psi_4x_3'r_V\phi_V(y-x_1)(\Phi^A-\Phi^V)$}\\\\	
		
		$H_{E_{c1}}^{SP}$
		&$\Psi_4(x_2'-1)x_3'r_{\Sigma}(-(\Phi^A-\Phi^V))\phi_V^T$
		&$\Psi_4(x_2'-1)x_3'r_{\Sigma}(-(\Phi^A-\Phi^V))\phi_V^t$
		&\tabincell{c}{$-2(1-x_2')((1-x_1)r_V\Phi^T\phi_V(\Psi_3^{-+}+\Psi_3^{+-})$\\$-\Psi_4x_3'r_{\Sigma}(\Phi^A-\Phi^V)\phi_V^t)$}\\\\
		
		$H_{E_{c2}}^{LL}$
		&\tabincell{c}{$\Psi_4(1-x_2')r_V(x_2+y-1)$\\$(\Phi^A-\Phi^V)(\phi_V^a+\phi_V^v)$}
		&$\Psi_4(1-x_2')r_V\phi_V(x_2+y-1)(\Phi^A-\Phi^V)	$
		&\tabincell{c}{$2r_V\phi_V(x_2+y-1)(\Phi^A-\Phi^V)$\\$(\Psi_2(x_2+x_3+y-1)-\Psi_4(x_2'-1))$}\\\\	
		
		$H_{E_{c2}}^{SP}$
		&\tabincell{c}{$\Psi_4(1-x_2')r_V(x_2+y-1)$\\$(\Phi^A+\Phi^V)(\phi_V^a+\phi_V^v)$}
		&$\Psi_4(x_2'-1)r_V\phi_V(x_2+y-1)(\Phi^A+\Phi^V)$
		&\tabincell{c}{$-2\Psi_4(1-x_2')r_V\phi_V$\\$(x_2+y-1)(\Phi^A+\Phi^V)$}\\\\
		
		$H_{E_{c3}}^{LL}$
		&\tabincell{c}{$x_1'(y-x_3)\phi_V^T(\Psi_2r_{\Sigma}(\Phi^A+\Phi^V)$\\$+\Psi_3 ^{-+}(r_{\Sigma}-1)\Phi^T)$}
		&\tabincell{c}{$x_1'(y-x_3)\phi_V^t(\Psi_2r_{\Sigma}(\Phi^A+\Phi^V)$\\$+\Psi_3 ^{-+}(r_{\Sigma}-1)\Phi^T)$}
		&\tabincell{c}{$-2(x_3-y)(\Psi_2r_V\phi_V(x_2+y-1)(\Phi^A-\Phi^V)$\\$-x_1'\phi_V^t(\Psi_2r_{\Sigma}(\Phi^A+\Phi^V)-\Psi_3 ^{-+}\Phi^T))$}\\\\	
		
		$H_{E_{c3}}^{SP}$
		&\tabincell{c}{$\Psi_4x_3'r_V(x_2+y-1)(\Phi^A+\Phi^V)(\phi_V^a$\\$+\phi_V^v)-\Psi_2x_1'r_{\Sigma}(y-x_3)(\Phi^A-\Phi^V)\phi_V^T$}
		&\tabincell{c}{$\Psi_4x_3'r_V\phi_V(x_2+y-1)(\Phi^A+\Phi^V)$\\$-\Psi_2x_1'r_{\Sigma}(y-x_3)(\Phi^A-\Phi^V)\phi_V^t$}
		&\tabincell{c}{$2r_V\phi_V(\Psi_4x_3'(x_2+y-1)(\Phi^A+\Phi^V)$\\$+\Psi_3 ^{-+}x_1'\Phi^T(y-x_3))-2\Psi_2x_1'r_{\Sigma}$\\$(y-x_3)(\Phi^A-\Phi^V)\phi_V^t$}\\		
		\hline\hline
		\end{tabular}
		\end{table}

\begin{table}[H]
	\footnotesize
	\centering
	TABLE~\ref{tab:amp} (continued)
	\newcommand{\tabincell}[2]{\begin{tabular}{@{}#1@{}}#2\end{tabular}}
	\begin{tabular}[t]{lccc}
		\hline\hline
		$$ &$\frac{A_1^T}{16M^4}$&$\frac{A_1^L}{16M^4}$&$\frac{A_2^L}{16M^4}$\\ \hline		
		$H_{E_{c4}}^{LL}$
		&\tabincell{c}{$-(1-x_2')(x_3+y-1)\phi_V^T(\Psi_2r_{\Sigma}$\\$(\Phi^A+\Phi^V)+\Psi_3 ^{-+}(r_{\Sigma}-1)\Phi^T)$}
		&\tabincell{c}{$-(1-x_2')(x_3+y-1)\phi_V^t(\Psi_2r_{\Sigma}$\\$(\Phi^A+\Phi^V)+\Psi_3 ^{-+}(r_{\Sigma}-1)\Phi^T)$}
		&\tabincell{c}{$2(x_2'-1)(x_3+y-1)(\Psi_2r_{\Sigma}(\Phi^A+\Phi^V)$\\$-\Psi_3 ^{-+}\Phi^T)\phi_V^t-2r_V\phi_V(y-x_1)$\\$(\Phi^A-\Phi^V)(\Psi_2(x_3+y-1)-\Psi_4x_3')$}\\\\	
		
		$H_{E_{c4}}^{SP}$
		&\tabincell{c}{$(1-x_2')(\Psi_4r_V(x_3+y-1)(\Phi^A+\Phi^V)$\\$(\phi_V^a+\phi_V^v)-r_{\Sigma}(\Phi^A-\Phi^V)\phi_V^T$\\$(\Psi_2(x_3+y-1)-\Psi_4x_3'))	$}
		&\tabincell{c}{$(1-x_2')(\Psi_4r_V\phi_V(x_3+y-1)$\\$(\Phi^A+\Phi^V)-r_{\Sigma}(\Phi^A-\Phi^V)\phi_V^t$\\$(\Psi_2(x_3+y-1)-\Psi_4x_3'))$}
		&\tabincell{c}{$2(1-x_2')(r_V\phi_V(x_3+y-1)((\Phi^A+\Phi^V)$\\$+\Psi_3 ^{-+}\Phi^T)\Psi_4-r_{\Sigma}(\Phi^A-\Phi^V)\phi_V^t$\\$(\Psi_2(x_3+y-1)-\Psi_4x_3'))$}\\\\
		
		$H_{E_{c5}}^{LL}$
		&\tabincell{c}{$(1-x_2')(x_3\Psi_4r_V(\Phi^A-\Phi^V)(\phi_V^a+\phi_V^v)$\\$-r_{\Sigma}(\Phi^A+\Phi^V)(\Psi_4(x_2'-1)$\\$+x_3\Psi_2)\phi_V^T+x_3\Psi_3^{+-}(r_{\Sigma}-1)\Phi^T\phi_V^T)$}
		&\tabincell{c}{$(1-x_2')(r_{\Sigma}(-\Phi^A-\Phi^V)(\Psi_4(x_2'-1)+x_3\Psi_2)$\\$\phi_V^t+x_3\Psi_4r_V\phi_V(\Phi^A-\Phi^V)$\\$+x_3\Psi_3^{+-}(r_{\Sigma}-1)\Phi^T\phi_V^t)$}
		&\tabincell{c}{$2(1-x_2')(x_3\Psi_4r_V\phi_V(\Phi^A-\Phi^V)$\\$-\phi_V^t(r_{\Sigma}(\Phi^A+\Phi^V)(\Psi_4(x_2'-1)$\\$+x_3\Psi_2)+x_3\Psi_3^{+-}\Phi^T))$}\\\\	
		
		$H_{E_{c5}}^{SP}$
		&$x_3\Psi_2(x_2'-1)r_{\Sigma}(\Phi^A-\Phi^V)\phi_V^T$
		&$x_3\Psi_2(x_2'-1)r_{\Sigma}(\Phi^A-\Phi^V)\phi_V^t$
		&\tabincell{c}{$-2x_3(1-x_2')(\Psi_2r_{\Sigma}(\Phi^A-\Phi^V)$\\$\phi_V^t+\Psi_3^{+-}r_V\Phi^T\phi_V)$}\\\\
		
		$H_{E_{c6}}^{LL}$
		&\tabincell{c}{$(1-x_2)x_1'\phi_V^T(\Psi_2r_{\Sigma}(\Phi^A+\Phi^V)$\\$+\Psi_3^{-+}(r_{\Sigma}-1)\Phi^T)-\Psi_4x_3'r_V$\\$(x_2+y-1)(\Phi^A-\Phi^V)(\phi_V^a+\phi_V^v)$}
		&\tabincell{c}{$(1-x_2)x_1'\phi_V^t(\Psi_2r_{\Sigma}(\Phi^A+\Phi^V)$\\$+\Psi_3^{-+}(r_{\Sigma}-1)\Phi^T)-\Psi_4x_3'r_V$\\$\phi_V(x_2+y-1)(\Phi^A-\Phi^V)$}
		&\tabincell{c}{$2((1-x_2)x_1'\phi_V^t(\Psi_2r_{\Sigma}(\Phi^A+\Phi^V)$\\$-\Psi_3^{-+}\Phi^T)+r_V\phi_V(x_2+y-1)$\\$(\Phi^A-\Phi^V)(\Psi_2-\Psi_4x_3'))$}\\\\	
		
		$H_{E_{c6}}^{SP}$
		&$(1-x_2)\Psi_2x_1'r_{\Sigma}(\Phi^A-\Phi^V)\phi_V^T$
		&$(1-x_2)\Psi_2x_1'r_{\Sigma}(\Phi^A-\Phi^V)\phi_V^t$
		&\tabincell{c}{$-2x_1'(r_V\phi_V(-x_2\Psi_4(\Phi^A+\Phi^V)$\\$-(x_2-1)\Psi_3^{-+}\Phi^T)+(x_2-1)$\\$\Psi_2r_{\Sigma}(\Phi^A-\Phi^V)\phi_V^t)$}\\\\
		
		$H_{E_{c7}}^{LL}$
		&$0$
		&$0$
		&$0$\\\\
		
		$H_{E_{c7}}^{SP}$
		&$0$
		&$0$
		&\tabincell{c}{$2x_3\Psi_4(x_2+x_3-1)r_V\phi_V(\Phi^A+\Phi^V)$}\\\\
		
		$H_{E_{d1}}^{LL}$
		&$\Psi_4x_3'r_V(\Phi^A-\Phi^V)(\phi_V^a+\phi_V^v)$
		&$\Psi_4x_3'r_V\phi_V(\Phi^A-\Phi^V)$
		&$2\Psi_4x_3'r_V\phi_V(\Phi^A-\Phi^V)$\\\\	
		
		$H_{E_{d1}}^{SP}$
		&$\Psi_4x_3'r_V(\Phi^A+\Phi^V)(\phi_V^a+\phi_V^v)$
		&$\Psi_4x_3'r_V\phi_V(\Phi^A+\Phi^V)$
		&\tabincell{c}{$2r_V\phi_V(\Psi_4x_3'(-(\Phi^A+\Phi^V))$\\$+(x_1-1)x_1'\Phi^T(\Psi_3^{-+}+\Psi_3^{+-}))$}\\\\
		
		$H_{E_{d2}}^{LL}$
		&\tabincell{c}{$(x_3-1)\Psi_2r_V(\Phi^V-\Phi^A)(\phi_V^a+\phi_V^v)$}
		&\tabincell{c}{$(x_3-1)\Psi_2r_V\phi_V(\Phi^V-\Phi^A)$}
		&\tabincell{c}{$-2(x_3-1)r_V\phi_V(\Phi^A-\Phi^V)(\Psi_2-\Psi_4x_1')$}\\\\	
		
		$H_{E_{d2}}^{SP}$
		&\tabincell{c}{$-r_V(\Phi^A+\Phi^V)(\Psi_4(x_2'-1)$\\$+(x_3-1)\Psi_2)(\phi_V^a+\phi_V^v)$}
		&\tabincell{c}{$-r_V\phi_V(\Phi^A+\Phi^V)$\\$(\Psi_4(x_2'-1)+(x_3-1)\Psi_2)$}
		&\tabincell{c}{$-2r_V\phi_V(\Psi_4(x_2'-1)(\Phi^A+\Phi^V)$\\$+(x_3-1)\Psi_2(\Phi^A+\Phi^V)$\\$-x_1'(1-x_3)\Psi_3 ^{+-}\Phi^T)$}\\\\
		
		$H_{E_{d3}}^{LL}$
		&\tabincell{c}{$(x_3-1)(\Psi_2r_V(x_3-y)(\Phi^A-\Phi^V)$\\$(\phi_V^a+\phi_V^v)+x_3'\Psi_3 ^{+-}$\\$(r_{\Sigma}-1)\Phi^T\phi_V^T)$}
		&\tabincell{c}{$(x_3-1)(\Psi_2r_V\phi_V(x_3-y)(\Phi^A-\Phi^V)$\\$+x_3'\Psi_3 ^{+-}(r_{\Sigma}-1)\Phi^T\phi_V^t)	$}
		&\tabincell{c}{$2(x_3-1)(\Psi_2r_V\phi_V(x_3-y)(\Phi^A-\Phi^V)$\\$-x_3'\Psi_3 ^{+-}\Phi^T\phi_V^t)$}\\\\	
		
		$H_{E_{d3}}^{SP}$
		&\tabincell{c}{$(x_3-1)\Psi_2r_V(x_3-y)(\Phi^A+\Phi^V)$\\$(\phi_V^a+\phi_V^v)+\Psi_4(x_2'-1)x_3'$\\$r_{\Sigma}(\Phi^A-\Phi^V)\phi_V^T$}
		&\tabincell{c}{$\Psi_4(x_2'-1)x_3'r_{\Sigma}(\Phi^A-\Phi^V)\phi_V^t$\\$+(x_3-1)\Psi_2r_V\phi_V(x_3-y)(\Phi^A+\Phi^V)$}
		&\tabincell{c}{$2r_V\phi_V(y-x_3)(\Psi_3 ^{-+}(x_2'-1)\Phi^T$\\$-(x_3-1)\Psi_2(\Phi^A+\Phi^V))+2\Psi_4(x_2'-1)$\\$x_3'r_{\Sigma}(\Phi^A-\Phi^V)\phi_V^t$}\\\\
		
		$H_{E_{d4}}^{LL}$
		&\tabincell{c}{$(x_3-1)x_3'\Psi_3 ^{+-}(1-r_{\Sigma})\Phi^T\phi_V^T$}
		&\tabincell{c}{$(x_3-1)x_3'\Psi_3 ^{+-}(1-r_{\Sigma})\Phi^T\phi_V^t$}
		&\tabincell{c}{$-2(x_3-1)x_3'(\Psi_4r_V\phi_V(\Phi^A-\Phi^V)$\\$-\Psi_3 ^{+-}\Phi^T\phi_V^t)$}\\\\	
		
		$H_{E_{d4}}^{SP}$
		&\tabincell{c}{$\Psi_4(1-x_2')(r_V(x_3+y-1)(\Phi^A+\Phi^V)$\\$(\phi_V^a+\phi_V^v)+x_3'r_{\Sigma}(\Phi^A-\Phi^V)\phi_V^T)$}
		&\tabincell{c}{$\Psi_4(1-x_2')(x_3'r_{\Sigma}(\Phi^A-\Phi^V)$\\$\phi_V^t+r_V\phi_V(x_3+y-1)(\Phi^A+\Phi^V))$}
		&\tabincell{c}{$2r_V\phi_V((x_3-1)x_3'\Psi_3 ^{+-}\Phi^T-(x_2'-1)$\\$(x_3+y-1)(\Psi_4(\Phi^A+\Phi^V)+\Psi_3 ^{-+}\Phi^T))$\\$-2\Psi_4(x_2'-1)x_3'r_{\Sigma}(\Phi^A-\Phi^V)\phi_V^t$}\\\\
		
		$H_{E_{d5}}^{LL}$
		&$0$
		&$0$
		&$0$\\\\	
		
		$H_{E_{d5}}^{SP}$
		&$0$
		&$0$
		&$0$\\
		\hline\hline
	\end{tabular}
\end{table}

\begin{table}[H]
	\footnotesize
	\centering
	TABLE~\ref{tab:amp} (continued)
	\newcommand{\tabincell}[2]{\begin{tabular}{@{}#1@{}}#2\end{tabular}}
	\begin{tabular}[t]{lccc}
		\hline\hline
		$$ &$\frac{A_1^T}{16M^4}$&$\frac{A_1^L}{16M^4}$&$\frac{A_2^L}{16M^4}$\\ \hline				
		$H_{E_{d6}}^{LL}$
		&\tabincell{c}{$r_V(\Phi^A-\Phi^V)((1-x_2)\Psi_2-\Psi_4x_3')(\phi_V^a+\phi_V^v)$}
		&\tabincell{c}{$r_V\phi_V(\Phi^A-\Phi^V)((1-x_2)\Psi_2-\Psi_4x_3')$}
		&\tabincell{c}{$-2r_V\phi_V(\Phi^A-\Phi^V)(\Psi_4x_3'+(x_2-1)\Psi_2)$}\\\\	
		
		$H_{E_{d6}}^{SP}$
		&\tabincell{c}{$(x_2-1)\Psi_2r_V(-(\Phi^A+\Phi^V))(\phi_V^a+\phi_V^v)$}
		&\tabincell{c}{$(x_2-1)\Psi_2r_V\phi_V(-(\Phi^A+\Phi^V))$}
		&\tabincell{c}{$2r_V\phi_V((x_2-1)\Psi_2(-(\Phi^A+\Phi^V))+x_1'$\\$(x_2\Psi_4(\Phi^A+\Phi^V)+(x_2-1)\Psi_3^{-+}\Phi^T))$}\\\\
		
		$H_{E_{d7}}^{LL}$
		&$0$
		&$0$
		&$-2x_3\Psi_4x_1'r_V\phi_V(\Phi^A-\Phi^V)$\\\\	
		
		$H_{E_{d7}}^{SP}$
		&$0$
		&$0$
		&$-2x_3\Psi_4x_1'r_V\phi_V(\Phi^A+\Phi^V)$\\\\
		
		$H_{E_{e1}}^{LL}$
		&\tabincell{c}{$-(1-x_2')\phi_V^T(r_{\Sigma}(-\Psi_4(x_2'-1)$\\$(\Phi^A+\Phi^V)-(y-1)\Phi^T(\Psi_3 ^{-+}+\Psi_3 ^{+-}))$\\$+(y-1)\Phi^T(\Psi_3 ^{-+}+\Psi_3 ^{+-}))$}
		&\tabincell{c}{$(x_2'-1)\phi_V^t(-\Psi_4(x_2'-1)r_{\Sigma}$\\$(\Phi^A+\Phi^V)-(y-1)(r_{\Sigma}-1)$\\$\Phi^T(\Psi_3 ^{-+}+\Psi_3 ^{+-}))$}
		&\tabincell{c}{$2(x_2'-1)\phi_V^t((y-1)\Phi^T(\Psi_3 ^{-+}+\Psi_3 ^{+-})$\\$-\Psi_4(x_2'-1)r_{\Sigma}(\Phi^A+\Phi^V))$\\$+2\Psi_2(y-1)r_V\phi_V(y-x_1)(\Phi^A-\Phi^V)$}\\\\	
		
		$H_{E_{e1}}^{SP}$
		&$\Psi_4(x_2'-1)^2r_{\Sigma}(-(\Phi^A-\Phi^V))\phi_V^T$
		&$\Psi_4(x_2'-1)^2r_{\Sigma}(-(\Phi^A-\Phi^V))\phi_V^t$
		&\tabincell{c}{$2(x_2'-1)((y-1)r_V\Phi^T\phi_V(\Psi_3 ^{-+}+\Psi_3 ^{+-})$\\$-\Psi_4(x_2'-1)r_{\Sigma}(\Phi^A-\Phi^V)\phi_V^t)$}\\\\
		
		$H_{E_{e2}}^{LL}$
		&$(x_3-1)\Psi_2(x_2'-1)r_{\Sigma}(-(\Phi^A+\Phi^V))\phi_V^T$
		&$(x_3-1)\Psi_2(x_2'-1)r_{\Sigma}(-(\Phi^A+\Phi^V))\phi_V^t$
		&$-2(x_3-1)\Psi_2(x_2'-1)r_{\Sigma}(\Phi^A+\Phi^V)\phi_V^t$\\\\	
		
		$H_{E_{e2}}^{SP}$
		&\tabincell{c}{$(x_2'-1)r_{\Sigma}(-(\Phi^A-\Phi^V))$\\$(\Psi_4(x_2'-1)+(x_3-1)\Psi_2)\phi_V^T$}
		&\tabincell{c}{$(x_2'-1)r_{\Sigma}(-(\Phi^A-\Phi^V))$\\$(\Psi_4(x_2'-1)+(x_3-1)\Psi_2)\phi_V^t$}
		&\tabincell{c}{$-2(1-x_2')((y-1)\Psi_3 ^{-+}r_V\Phi^T\phi_V$\\$-r_{\Sigma}(\Phi^A-\Phi^V)(\Psi_4(x_2'-1)$\\$+(x_3-1)\Psi_2)\phi_V^t)$}\\\\
		
		$H_{E_{e3}}^{LL}$
		&\tabincell{c}{$\phi_V^T((x_2'-1)r_{\Sigma}(\Phi^A+\Phi^V)(\Psi_4(x_2'-1)$\\$+\Psi_2(y-x_2))-(y-1)(r_{\Sigma}-1)$\\$\Phi^T(\Psi_3 ^{+-}-\Psi_3 ^{-+}(x_2'-1)))$}
		&\tabincell{c}{$\phi_V^t((x_2'-1)r_{\Sigma}(\Phi^A+\Phi^V)(\Psi_4(x_2'-1)$\\$+\Psi_2(y-x_2))-(y-1)(r_{\Sigma}-1)$\\$\Phi^T(\Psi_3 ^{+-}-\Psi_3 ^{-+}(x_2'-1)))$}
		&\tabincell{c}{$-2(\phi_V^t((y-1)\Phi^T(\Psi_3 ^{-+}(x_2'-1)-\Psi_3 ^{+-})$\\$-(x_2'-1)r_{\Sigma}(\Phi^A+\Phi^V)(\Psi_4(x_2'-1)$\\$+\Psi_2(y-x_2)))+(y-1)r_V\phi_V$\\$(\Phi^A-\Phi^V)(\Psi_4x_2'+\Psi_2(1-x_2+y)))$}\\\\	
		
		$H_{E_{e3}}^{SP}$
		&\tabincell{c}{$\Psi_2(x_2'-1)r_{\Sigma}(y-x_2)(\Phi^A-\Phi^V)\phi_V^T$\\$-(y-1)\Psi_3 ^{-+}r_V\Phi^T(y-x_2-1)(\phi_V^a-\phi_V^v)$}
		&\tabincell{c}{$\Psi_2(x_2'-1)r_{\Sigma}(y-x_2)(\Phi^A-\Phi^V)\phi_V^t$\\$+(y-1)\Psi_3 ^{-+}r_V\Phi^T\phi_V(1-x_2+y)$}
		&\tabincell{c}{$2\Psi_2(x_2'-1)r_{\Sigma}(y-x_2)(\Phi^A-\Phi^V)$\\$\phi_V^t+2(y-1)r_V\Phi^T\phi_V$\\$(\Psi_3 ^{-+}(y-x_2-1)+\Psi_3 ^{+-})$}\\\\
		
		$H_{E_{e4}}^{LL}$
		&\tabincell{c}{$(1-y)\Psi_3 ^{+-}(r_{\Sigma}-1)\Phi^T\phi_V^T$}
		&\tabincell{c}{$(1-y)\Psi_3 ^{+-}(r_{\Sigma}-1)\Phi^T\phi_V^t$}
		&\tabincell{c}{$-2(y-1)(\Psi_4r_V\phi_V(\Phi^A-\Phi^V)$\\$-\Psi_3 ^{+-}\Phi^T\phi_V^t)$}\\\\	
		
		$H_{E_{e4}}^{SP}$
		&\tabincell{c}{$(1-y)\Psi_3 ^{-+}r_V\Phi^T(x_3+y-1)(\phi_V^a-\phi_V^v)$}
		&\tabincell{c}{$(y-1)\Psi_3 ^{-+}r_V\Phi^T\phi_V(x_3+y-1)$}
		&\tabincell{c}{$2(y-1)r_V\Phi^T\phi_V(\Psi_3 ^{-+}$\\$(x_3+y-1)+\Psi_3 ^{+-})$}\\\\
		
		$H_{E_{f1}}^{LL}$
		&\tabincell{c}{$\Psi_4(x_2'-1)r_V(x_3+y-1)$\\$(\Phi^V-\Phi^A))(\phi_V^a+\phi_V^v)$}
		&\tabincell{c}{$\Psi_4(x_2'-1)r_V\phi_V(x_3+y-1)(\Phi^V-\Phi^A)$}
		&\tabincell{c}{$-2\Psi_4r_V\phi_V(\Phi^A-\Phi^V)$\\$(x_3'(1-y-x_3)+x_2x_1')$}\\\\	
		
		$H_{E_{f1}}^{SP}$
		&\tabincell{c}{$\Psi_4(1-x_2')r_V(x_3+y-1)$\\$(\Phi^A+\Phi^V)(\phi_V^a+\phi_V^v)$}
		&\tabincell{c}{$\Psi_4(1-x_2')r_V\phi_V(x_3+y-1)(\Phi^A+\Phi^V)$}
		&\tabincell{c}{$-2\Psi_4(x_2'-1)r_V\phi_V$\\$(x_3+y-1)(\Phi^A+\Phi^V)$}\\\\
		
		$H_{E_{f2}}^{LL}$
		&\tabincell{c}{$-(x_3-1)x_1'\phi_V^T(\Psi_2r_{\Sigma}(\Phi^A+\Phi^V)$\\$-\Psi_3 ^{+-}(r_{\Sigma}-1)\Phi^T)$}
		&\tabincell{c}{$-(x_3-1)x_1'\phi_V^t(\Psi_2r_{\Sigma}(\Phi^A+\Phi^V)$\\$-\Psi_3 ^{+-}(r_{\Sigma}-1)\Phi^T)$}
		&\tabincell{c}{$2(x_3-1)x_1'(\Psi_4r_V\phi_V(\Phi^A-\Phi^V)$\\$-\phi_V^t(\Psi_2r_{\Sigma}(\Phi^A+\Phi^V)+\Psi_3 ^{+-}\Phi^T))$}\\\\	
		
		$H_{E_{f2}}^{SP}$
		&\tabincell{c}{$(1-x_3)\Psi_2x_1'r_{\Sigma}(\Phi^A-\Phi^V)\phi_V^T$\\$-\Psi_4(x_2'-1)r_V(x_3+y-1)$\\$(\Phi^A+\Phi^V)(\phi_V^a+\phi_V^v)$}
		&\tabincell{c}{$(1-x_3)\Psi_2x_1'r_{\Sigma}(\Phi^A-\Phi^V)\phi_V^t$\\$-\Psi_4(x_2'-1)r_V\phi_V$\\$(x_3+y-1)(\Phi^A+\Phi^V)$}
		&\tabincell{c}{$2(1-x_3)\Psi_2x_1'r_{\Sigma}(\Phi^A-\Phi^V)\phi_V^t$\\$+2r_V\phi_V((1-x_3)x_1'\Psi_3 ^{+-}\Phi^T-\Psi_4$\\$(x_2'-1)(x_3+y-1)(\Phi^A+\Phi^V))$}\\\\
		
		$H_{E_{f3}}^{LL}$
		&\tabincell{c}{$\Psi_4(x_2'-1)r_V(x_3+y-1)(\Phi^A-\Phi^V)$\\$(\phi_V^a+\phi_V^v)+x_1'(x_2-y)\phi_V^T(\Psi_2r_{\Sigma}$\\$(\Phi^A+\Phi^V)-\Psi_3^{+-}(r_{\Sigma}-1)\Phi^T)$}
		&\tabincell{c}{$x_1'(x_2-y)\phi_V^t(\Psi_2r_{\Sigma}(\Phi^A+\Phi^V)$\\$-\Psi_3^{+-}(r_{\Sigma}-1)\Phi^T)+\Psi_4(x_2'-1)r_V$\\$\phi_V(x_3+y-1)(\Phi^A-\Phi^V)$}
		&\tabincell{c}{$2x_1'(x_2-y)\phi_V^t(\Psi_2r_{\Sigma}(\Phi^A+\Phi^V)$\\$+\Psi_3^{+-}\Phi^T)-2\Psi_4r_V\phi_V$\\$(\Phi^A-\Phi^V)((1-x_2')(x_3+y)-x_3'$}\\\\	
		
		$H_{E_{f3}}^{SP}$
		&\tabincell{c}{$\Psi_2x_1'r_{\Sigma}(x_2-y)(-(\Phi^A-\Phi^V))\phi_V^T$}
		&\tabincell{c}{$\Psi_2x_1'r_{\Sigma}(x_2-y)(-(\Phi^A-\Phi^V))\phi_V^t$}
		&\tabincell{c}{$2r_V\phi_V(\Psi_2(y-x_2-1)(x_3+y-1)$\\$(\Phi^A+\Phi^V)-x_1'\Psi_3^{+-}\Phi^T(x_2-y))$\\$-2\Psi_2x_1'r_{\Sigma}(x_2-y)(\Phi^A-\Phi^V)\phi_V^t$}\\
		\hline\hline
	\end{tabular}
\end{table}

\begin{table}[H]
	\footnotesize
	\centering
	TABLE~\ref{tab:amp} (continued)
	\newcommand{\tabincell}[2]{\begin{tabular}{@{}#1@{}}#2\end{tabular}}
	\begin{tabular}[t]{lccc}
		\hline\hline
		$$ &$\frac{A_1^T}{16M^4}$&$\frac{A_1^L}{16M^4}$&$\frac{A_2^L}{16M^4}$\\ \hline				
		$H_{E_{f4}}^{LL}$
		&$0$
		&$0$
		&$0$\\\\	
		
		$H_{E_{f4}}^{SP}$
		&$0$
		&$0$
		&$2\Psi_2r_V\phi_V(x_3+y-1)^2(\Phi^A+\Phi^V)$\\\\
		
		$H_{E_{g1}}^{LL}$
		&\tabincell{c}{$\frac{1}{2}(x_2'-1)(\Psi_4r_V(2x_3+y-1)(\Phi^A-\Phi^V)$\\$(\phi_V^a+\phi_V^v)+r_{\Sigma}(-(\Phi^A+\Phi^V))\phi_V^T$\\$(\Psi_2(2x_3+y-1)-x_1'\Psi_4)-(r_{\Sigma}-1)\Phi^T$\\$\phi_V^T(\Psi_3 ^{-+}(x_3+2y-2)+\Psi_3^{+-}(1-x_3+y)))$}
		&\tabincell{c}{$\frac{1}{2}(x_2'-1)(r_{\Sigma}(-(\Phi^A+\Phi^V))\phi_V^t(\Psi_2$\\$(2x_3+y-1)-x_1'\Psi_4)+\Psi_4r_V\phi_V$\\$(2x_3+y-1)(\Phi^A-\Phi^V)-(r_{\Sigma}-1)\Phi^T\phi_V^t$\\$(\Psi_3 ^{-+}(x_3+2y-2)+\Psi_3^{+-}(1-x_3+y)))$}
		&\tabincell{c}{$r_V\phi_V(\Phi^A-\Phi^V)(\Psi_4(2x_3'(x_1-y)$\\$-(x_2-x_3)(x_2'-1)) +\Psi_2(x_3+2y-2)$\\$(y-x_1))-(x_2'-1)\phi_V^t(r_{\Sigma}(\Phi^A+\Phi^V)$\\$(\Psi_2(2x_3+y-1)-x_1'\Psi_4)+\Phi^T(\Psi_3^{+-}$\\$(x_3-y+1)-\Psi_3 ^{-+}(x_3+2y-2)))$}\\\\	
		
		$H_{E_{g1}}^{SP}$
		&\tabincell{c}{$\frac{1}{2}(1-x_2')(\Psi_4r_V(2x_3+y-1)(\Phi^A+\Phi^V)$\\$(\phi_V^a+\phi_V^v)-r_{\Sigma}(\Phi^A-\Phi^V)\phi_V^T$\\$(\Psi_4(x_2'-x_3'-1)+\Psi_2(2x_3+y-1)))$}
		&\tabincell{c}{$\frac{1}{2}(1-x_2')(\Psi_4r_V\phi_V(2x_3+y-1)$\\$(\Phi^A+\Phi^V)-r_{\Sigma}(\Phi^A-\Phi^V)\phi_V^t$\\$(\Psi_2(2x_3+y-1)-x_1'\Psi_4))$}
		&\tabincell{c}{$(1-x_2')(r_V\phi_V(\Psi_4(2x_3+y-1)$\\$(\Phi^A+\Phi^V)+\Phi^T(\Psi_3 ^{-+}(x_3+2y-2)$\\$+\Psi_3^{+-}(1-x_3+y)))-r_{\Sigma}(\Phi^A-\Phi^V)$\\$\phi_V^t(\Psi_2(2x_3+y-1)-x_1'\Psi_4))$}\\\\
		
		$H_{E_{g2}}^{LL}$
		&\tabincell{c}{$\frac{1}{2}(x_3-1)(\Psi_2r_V(x_3-y+1)(\Phi^A-\Phi^V)$\\$(\phi_V^a+\phi_V^v)+\Psi_2(2x_2'+x_3'-2)$\\$r_{\Sigma}(-(\Phi^A+\Phi^V))\phi_V^T+(x_3'-x_1')$\\$\Psi_3 ^{+-}(r_{\Sigma}-1)\Phi^T\phi_V^T)$}
		&\tabincell{c}{$\frac{1}{2}(x_3-1)(\Psi_2(2x_2'+x_3'-2)r_{\Sigma}$\\$(-(\Phi^A+\Phi^V))\phi_V^t+\Psi_2r_V\phi_V$\\$(x_3-y+1)(\Phi^A-\Phi^V)+(x_3'-x_1')$\\$\Psi_3 ^{+-}(r_{\Sigma}-1)\Phi^T\phi_V^t)$}
		&\tabincell{c}{$(x_3-1)(\phi_V^t(-\Psi_2(2x_2'+x_3'-2)$\\$r_{\Sigma}(\Phi^A+\Phi^V)-(x_3'-x_1')\Psi_3 ^{+-}\Phi^T)$\\$+r_V\phi_V(\Phi^A-\Phi^V)(\Psi_4(x_3'-x_1')$\\$+\Psi_2(x_3-y+1)))$}\\\\
		
		$H_{E_{g2}}^{SP}$
		&\tabincell{c}{$\frac{1}{2}(r_{\Sigma}(\Phi^A-\Phi^V)((x_3-1)\Psi_2$\\$(2x_2'+x_3'-2)+\Psi_4(1-x_2')x_1')\phi_V^T$\\$-r_V(\Phi^A+\Phi^V)(\phi_V^a+\phi_V^v)(\Psi_4(x_2'-1)$\\$(2x_3+y-1)+(x_3-1)\Psi_2(x_3-y+1)))$}
		&\tabincell{c}{$\frac{1}{2}(r_{\Sigma}(\Phi^A-\Phi^V)((x_3-1)\Psi_2$\\$(2x_2'+x_3'-2)+\Psi_4(1-x_2')x_1')\phi_V^t$\\$-r_V\phi_V(\Phi^A+\Phi^V)(\Psi_4(2x_3+y-1)$\\$(x_2'-1)+(x_3-1)\Psi_2(x_3-y+1)))$}
		&\tabincell{c}{$r_{\Sigma}(\Phi^A-\Phi^V)((x_3-1)\Psi_2(2x_2'+x_3'-2)$\\$+\Psi_4(1-x_2')x_1')\phi_V^t-r_V\phi_V((\Phi^A+\Phi^V)$\\$(\Psi_4(x_2'-1)(2x_3+y-1)+(x_3-1)\Psi_2$\\$(x_3-y+1))+\Phi^T(\Psi_3 ^{-+}(x_2'-1)$\\$(x_3+2y-2)-(x_3-1)(x_3'-x_1')\Psi_3 ^{+-}))$}\\\\	
		
		$H_{E_{g3}}^{LL}$
		&\tabincell{c}{$\frac{1}{2}(r_V(\Phi^A-\Phi^V)(\phi_V^a+\phi_V^v)(\Psi_2(x_2-y)$\\$(x_3-y+1)-\Psi_4(x_2'-1)(2x_3+y-1))$\\$+\phi_V^T(r_{\Sigma}((\Phi^A+\Phi^V)(\Psi_4x_1'(1-x_2')$\\$+\Psi_2((x_2'-x_1'-1)(y-x_2)-2x_3-y+1))$\\$+\Phi^T(\Psi_3 ^{-+}(x_2'-1)(x_3+2y-2)+\Psi_3 ^{+-}$\\$(-yx_2'+2x_3'(x_2-y)+x_2x_2'2x_3-x_1)))$\\$+\Phi^T(\Psi_3 ^{+-}((x_2'+2x_3'-1)(y-x_2)-x_3$\\$+y-1)-\Psi_3 ^{-+}(x_2'-1)(x_3+2y-2))))$}
		&\tabincell{c}{$\frac{1}{2}(\Phi^Ar_{\Sigma}\phi_V^t(\Psi_4x_1'(1-x_2')+\Psi_2((2x_2'$\\$+x_3'-2)(y-x_2)-2x_3-y+1))+r_V$\\$\phi_V(\Phi^A-\Phi^V)(\Psi_2(x_2-y)(x_3-y+1)$\\$-\Psi_4(x_2'-1)(2x_3+y-1))-(r_{\Sigma}-1)$\\$\Phi^T\phi_V^t(\Psi_3 ^{+-}((x_2'+2x_3'-1)(y-x_2)-x_3$\\$+y-1)-\Psi_3 ^{-+}(x_2'-1)(x_3+2y-2))$\\$+r_{\Sigma}\Phi^V\phi_V^t(\Psi_4x_1'(1-x_2')+\Psi_2((2x_2'$\\$+x_3'-2)(y-x_2)-2x_3-y+1)))$}
		&\tabincell{c}{$\phi_V^t(r_{\Sigma}(\Phi^A+\Phi^V)(\Psi_4x_1'(1-x_2')+\Psi_2$\\$((x_2'-x_1'-1)(y-x_2)-2x_3-y+1))$\\$+\Phi^T(\Psi_3 ^{+-}((x_2'+2x_3'-1)(y-x_2)$\\$-x_3+y-1)-\Psi_3 ^{-+}(x_2'-1)$\\$(x_3+2y-2)))+r_V\phi_V(\Phi^A-\Phi^V)(\Psi_2$\\$(x_3(2x_2-2y+1)-(y-1)(y-x_2-2))$\\$-\Psi_4(x_2'(x_3+2y-1)+2x_3'-2x_3-y))$}\\\\
		
		$H_{E_{g3}}^{SP}$
		&\tabincell{c}{$\frac{1}{2}(\Psi_2r_{\Sigma}(-(\Phi^A-\Phi^V))((x_2'-x_1'-1)$\\$(y-x_2)-2x_3-y+1)\phi_V^T-r_V$\\$(\Psi_2(y-x_2)(-x_3+y-1)(\Phi^A+\Phi^V)$\\$(\phi_V^a+\phi_V^v)-\Psi_3 ^{-+}\Phi^T(y-x_2-1)$\\$(x_3+2y-2)(\phi_V^a-\phi_V^v)))$}
		&\tabincell{c}{$\frac{1}{2}(\Psi_2r_{\Sigma}(-(\Phi^A-\Phi^V))((x_2'-x_1'-1)$\\$(y-x_2)-2x_3-y+1)\phi_V^t-r_V$\\$\phi_V(\Psi_2(x_2-y)(x_3-y+1)(\Phi^A+\Phi^V)$\\$+\Psi_3 ^{-+}\Phi^T(y-x_2-1)(x_3+2y-2)))$}
		&\tabincell{c}{$\Psi_2r_{\Sigma}(-(\Phi^A-\Phi^V))((2x_2'+x_3'-2)$\\$(y-x_2)-2x_3-y+1)\phi_V^t-r_V\phi_V(\Psi_2$\\$(y(y-2x_3)+x_2(2x_3+y-1)+x_3+3y$\\$-2)(\Phi^A+\Phi^V)-\Phi^T(\Psi_3 ^{-+}(x_2-y+1)$\\$(x_3+2y-2)+\Psi_3 ^{+-}(2x_3'(x_2-y)$\\$-yx_2'+x_2x_2'-x_2+x_3+1)))$}\\\\	
		
		$H_{E_{g4}}^{LL}$
		&\tabincell{c}{$\frac{1}{4}r_V(1-2x_3-y)(\Phi^A+\Phi^V)(\phi_V^a+\phi_V^v)$\\$(\Psi_3^{+-}+4\Psi_4r_V\Phi^T(x_3-y+1)(\phi_V^a-\phi_V^v)$\\$+r_{\Sigma}\phi_V^T(\Psi_3^{+-}(x_2'(\Phi^A-\Phi^V)-2x_3\Phi^T$\\$+x_3'(\Phi^V-\Phi^A)+2(y-1)\Phi^T+\Phi^V)$\\$+2\Psi_2(2x_3+y-1)(\Phi^A+\Phi^V))$\\$-\Phi^A+2\Psi_3^{+-}\Phi^T(x_3-y+1)\phi_V^T)$}
		&\tabincell{c}{$\frac{1}{4}(r_{\Sigma}\phi_V^t(\Psi_3^{+-}(x_2'(\Phi^A-\Phi^V)$\\$+x_3'(\Phi^V-\Phi^A)-\Phi^A-2x_3\Phi^T$\\$+2(y-1)\Phi^T+\Phi^V)+2\Psi_2(2x_3+y-1)$\\$(\Phi^A+\Phi^V))+r_V\phi_V(4\Psi_4\Phi^T(y-x_3-1)$\\$-\Psi_3^{+-}(2x_3+y-1)(\Phi^A+\Phi^V))$\\$-2\Psi_3^{+-}\Phi^T(y-x_3-1)\phi_V^t)$}
		&\tabincell{c}{$\frac{1}{2}(\phi_V^t(r_{\Sigma}(x_1'\Psi_3^{+-}(\Phi^V-\Phi^A)$\\$+2\Psi_2(2x_3+y-1)(\Phi^A+\Phi^V))$\\$-2\Psi_3^{+-}\Phi^T(y-x_3-1))-r_V\phi_V$\\$(-(\Phi^A+\Phi^V)(\Psi_3^{-+}(x_3+2y-2)-\Psi_3^{+-}$\\$(2x_3+y-1))-2\Psi_4(y-x_3-1)(\Phi^A$\\$+2\Phi^T-\Phi^V)+4\Psi_2\Phi^T(x_3+2y-2)))$}\\\\
		
		$H_{E_{g4}}^{SP}$
		&\tabincell{c}{$\frac{1}{8}(\Phi^A\Psi_3^{+-}r_V(2x_3+y-1)(\phi_V^a+\phi_V^v)$\\$-2\Psi_3^{-+}r_V\Phi^T(x_3+y-1)(x_3+2y-2)$\\$(\phi_V^a-\phi_V^v)-\Psi_3^{+-}r_V\Phi^V(2x_3+y-1)$\\$(\phi_V^a+\phi_V^v)-r_{\Sigma}(2\Psi_2(2x_3(\Phi^A+\Phi^T-\Phi^V)$\\$\phi_V^T+(y-1)(\Phi^A+4\Phi^T-\Phi^V))-x_1'\Psi_3^{+-}$\\$(\Phi^A+\Phi^V))+4\Psi_2\Phi^(x_3+2y-2)\phi_V^T)$}
		&\tabincell{c}{$\frac{1}{8}(r_V\phi_V(\Psi_3^{+-}(2x_3+y-1)(\Phi^A-\Phi^V)$\\$+2\Psi_3^{-+}\Phi^T(x_3+y-1)(x_3+2y-2))$\\$+\phi_V^t(2\Psi_2(r_{\Sigma}((y-1)(-\Phi^A-4\Phi^T+\Phi^V)$\\$-2x_3(\Phi^A+\Phi^T-\Phi^V))+2\Phi^T$\\$(x_3+2y-2))+x_1'\Psi_3^{+-}r_{\Sigma}(\Phi^A+\Phi^V)))$}
		&\tabincell{c}{$\frac{1}{4}(\phi_V^t(-(2\Psi_2(r_{\Sigma}(2x_3+y-1)$\\$(\Phi^A-\Phi^V)-2\Phi^T(x_3+2y-2))$\\$-x_1'\Psi_3^{+-}r_{\Sigma}(\Phi^A+\Phi^V)))-r_V\phi_V$\\$((x_3+2y-2)(2\Psi_2(x_3+y-1)$\\$(\Phi^A+\Phi^V)-\Psi_3^{+-}(\Phi^A-\Phi^V))$\\$-2\Phi^T(\Psi_3^{-+}(x_3+y-1)(x_3+2y-2)$\\$+\Psi_3^{+-}(y-x_3-1))))$}\\		
		\hline\hline
	\end{tabular}
\end{table}

\begin{table}[H]
	\footnotesize
	\centering
	TABLE~\ref{tab:amp} (continued)
	\newcommand{\tabincell}[2]{\begin{tabular}{@{}#1@{}}#2\end{tabular}}
	\begin{tabular}[t]{lccc}
		\hline\hline
		$$ &$\frac{A_1^T}{16M^4}$&$\frac{A_1^L}{16M^4}$&$\frac{A_2^L}{16M^4}$\\ \hline				
		$H_{B_{a1}}^{LL}$
		&$0$
		&$0$
		&$0$\\\\
		
		$H_{B_{a1}}^{SP}$
		&\tabincell{c}{$r_V(\phi_V^a(-\Psi_4(\Phi^A+\Phi^V)(x_3-x_2x_3'+y-1)$\\$-(x_2+1)\Psi_3 ^{-+}\Phi^T(x_3+y-1))$\\$+\phi_V^v((x_2+1)\Psi_3 ^{-+}\Phi^T(x_3+y-1)$\\$-\Psi_4(\Phi^A+\Phi^V)(x_2x_3'+x_3+y-1)))$}
		&\tabincell{c}{$-r_V\phi_V(\Psi_4(\Phi^A+\Phi^V)$\\$(x_2x_3'+x_3+y-1)-(x_2+1)$\\$\Psi_3 ^{-+}\Phi^T(x_3+y-1))$}
		&\tabincell{c}{$-2r_V\phi_V(-\Psi_4(\Phi^A+\Phi^V)$\\$(x_3'-x_3-y+1)-(x_2-1)$\\$\Psi_3 ^{-+}\Phi^T(x_3+y-1))$}\\\\	
		
		$H_{B_{a2}}^{LL}$
		&\tabincell{c}{$(x_3-1)\Psi_2r_V(x_3+y-1)$\\$(-(\Phi^A-\Phi^V))(\phi_V^a-\phi_V^v)$}
		&\tabincell{c}{$(x_3-1)\Psi_2r_V\phi_V$\\$(x_3+y-1)(\Phi^A-\Phi^V)$}
		&\tabincell{c}{$2(x_3-1)r_V\phi_V(\Phi^A-\Phi^V)$\\$(\Psi_2(x_3+y-1)-\Psi_4x_3')$}\\\\
		
		$H_{B_{a2}}^{SP}$
		&\tabincell{c}{$r_V(x_3+y-1)(\phi_V^a-\phi_V^v)(\Psi_4(x_1'-1)$\\$(\Phi^A+\Phi^V)+(x_3-1)\Psi_3 ^{-+}\Phi^T)	$}
		&\tabincell{c}{$-r_V\phi_V(x_3+y-1)$\\$(\Psi_4(1-x_1')(\Phi^A+\Phi^V)$\\$+(x_3-1)\Psi_3 ^{-+}\Phi^T)$}
		&\tabincell{c}{$-2r_V\phi_V(x_2'(x_3+y-1)$\\$(\Psi_4(\Phi^A+\Phi^V)+2\Psi_3 ^{-+}\Phi^T)$\\$+x_3'(\Psi_4(2x_3+y-2)(\Phi^A+\Phi^V)$\\$+(x_3+1)\Psi_3 ^{-+}\Phi^T(x_3+y-1)))$}\\\\	
		
		$H_{B_{a3}}^{LL}$
		&$0$
		&$0$
		&$0$\\\\
		
		$H_{B_{a3}}^{SP}$
		&\tabincell{c}{$-x_1r_V(\Phi^A+\Phi^V)(\phi_V^a-\phi_V^v)$\\$(\Psi_2(x_3+y-1)-\Psi_4x_3')$}
		&\tabincell{c}{$-x_1r_V\phi_V(-(\Phi^A+\Phi^V))$\\$(\Psi_2(x_3+y-1)-\Psi_4x_3')$}
		&\tabincell{c}{$2x_1r_V\phi_V(x_3'(\Psi_3 ^{+-}\Phi^T-\Psi_4(\Phi^A+\Phi^V))$\\$+\Psi_2(x_3+y-1)(\Phi^A+\Phi^V))$}\\\\		
		
		$H_{B_{a4}}^{LL}$
		&\tabincell{c}{$(x_3-1)\Psi_2r_V(x_3+y-1)$\\$(\Phi^A-\Phi^V)(\phi_V^a-\phi_V^v)$}
		&\tabincell{c}{$(x_3-1)\Psi_2r_V\phi_V(1-x_3-y)(\Phi^A-\Phi^V)$}
		&\tabincell{c}{$-2(x_3-1)r_V\phi_V(\Phi^A-\Phi^V)$\\$(\Psi_2(x_3+y-1)-\Psi_4x_3')$}\\\\
		
		$H_{B_{a4}}^{SP}$
		&\tabincell{c}{$r_V(x_3+y-1)(\Phi^A+\Phi^V)(\Psi_4(x_2'-1)$\\$(\phi_V^a+\phi_V^v)+(x_3-1)\Psi_2(\phi_V^a-\phi_V^v))$}
		&\tabincell{c}{$r_V\phi_V(x_3+y-1)(-(\Phi^A+\Phi^V))$\\$((x_3-1)\Psi_2-\Psi_4(x_2'-1))$}
		&\tabincell{c}{$-2r_V\phi_V((1-x_2')(x_3+y-1)$\\$(\Psi_4(\Phi^A+\Phi^V)+2\Psi_3^{-+}\Phi^T)$\\$+(x_3-1)\Psi_2(x_3+y-1)(\Phi^A+\Phi^V)$\\$+(x_3-1)x_3'\Psi_3^{+-}\Phi^T)$}\\\\		
		\hline\hline
		\end{tabular}
		\end{table}

\begin{table}[H]
	\tiny
	\centering
	\caption{The expressions of $H^{\sigma}_{R_{ij}}$ in the invariant amplitudes $A_1^{L}$ and $A_2^L$ for the $\Lambda_b\rightarrow \Sigma^0J/\psi$ decay.}
	\newcommand{\tabincell}[2]{\begin{tabular}{@{}#1@{}}#2\end{tabular}}
	\label{tab:ampj}
	\begin{tabular}[t]{lcc}
		\hline\hline
		$$ &$\frac{A_1^L}{16M^4}$&$\frac{A_2^L}{16M^4}$\\ \hline
		$H_{B_{a1}}^{LL}$
		&\tabincell{c}{$\frac{3}{2}r_{\Sigma}\Phi^T\{r_c\psi^t(\Psi_3^{-+}x_3'+\Psi_3^{+-})$\\$+\psi^Lr_V[x_2x_3'\Psi_3^{+-}-(x_3+y-1)(\Psi_3^{-+}x_3'-\Psi_3^{+-})]\}$\\$+r_{\Sigma}(\Phi^A-\Phi^V)\{r_c\psi^t[\Psi_4(x_3'-1)+(x_2+1)\Psi_2]$\\$+\psi^Lr_V[\Psi_4x_3'(x_3+y)-(x_2\Psi_2+\Psi_2+\Psi_4)(x_3+y-1)]\}$}
		&\tabincell{c}{$3r_{\Sigma}\Phi^T\{r_c\psi^t(\Psi_3^{-+}x_3'+\Psi_3^{+-})$\\$+\psi^Lr_V[\Psi_3^{-+}(x_3'-2)(x_3+y-1)+\Psi_3^{+-}(x_3-x_3'+y-1)]\}$\\$+2r_{\Sigma}(\Phi^A-\Phi^V)(r_c\psi^t(\Psi_4(x_3'-1)+(x_2+1)\Psi_2)$\\$+\psi^Lr_V(\Psi_4(x_3'-1)(x_3-x_3'+y-1)-(x_2-1)\Psi_2(x_3+y-1)))$}\\\\
		
		$H_{B_{a1}}^{SP}$
		&\tabincell{c}{$\frac{1}{2}\Psi_3^{-+}\Phi^T\{\psi^Lr_V[x_3'r_{\Sigma}(y-x_1)-(x_3+y-1)$\\$(r_{\Sigma}+2x_2r_{\Sigma}-2x_2-2)]-r_c\psi^t[(x_3'-2x_2-3)r_{\Sigma}+2(x_2+1)]\}$\\$+(\Phi^A+\Phi^V)\{r_c\psi^t[\Psi_4(r_{\Sigma}-1)+(x_2+1)\Psi_2r_{\Sigma}]$\\$+\psi^Lr_V[\Psi_4(r_{\Sigma}-1)(x_2x_3'+x_3+y-1)-(x_2+1)\Psi_2r_{\Sigma}(x_3+y-1)]\}$}
		&\tabincell{c}{$\Psi_3^{-+}\Phi^T\{r_c\psi^t[(1-x_3')r_{\Sigma}-2(x_2+1)]$\\$-\psi^Lr_V[x_3'r_{\Sigma}(x_3+y)+(x_3+y-1)(r_{\Sigma}-2x_2+2)]\}$\\$+2(\Phi^A+\Phi^V)\{r_c\psi^t[(x_2+1)\Psi_2r_{\Sigma}-\Psi_4]$\\$+\psi^Lr_V[\Psi_4x_3'(r_{\Sigma}x_2+r_{\Sigma}+1)-(x_3+y-1)(x_2\Psi_2r_{\Sigma}-\Psi_2r_{\Sigma}+\Psi_4)]\}$}\\\\
				
		$H_{B_{a2}}^{LL}$
		&\tabincell{c}{$\Psi_3^{+-}r_{\Sigma}\Phi^T\{r_c(1-x_1')\psi^t+\psi^Lr_V[x_2'(x_3+y-1)+x_3'(2x_3+y-2)]\}$\\$+(\Phi^V-\Phi^A)\{\psi^Lr_V(x_3+y-1)[(x_3-1)\Psi_2(r_{\Sigma}-1)-\Psi_4(x_1'-1)r_{\Sigma}]$\\$-r_c\psi^t[\Psi_4(x_1'-1)r_{\Sigma}+(x_3-1)\Psi_2(r_{\Sigma}-1)]\}$}
		&\tabincell{c}{$r_{\Sigma}\Phi^T\langle2r_c(1-x_1')\Psi_3^{+-}\psi^t+\psi^Lr_V\{x_3'[\Psi_3^{-+}(5-5x_3-4y)$\\$+2\Psi_3^{+-}(2x_3+y-2)]-2x_2'(x_3+y-1)(2\Psi_3^{-+}-\Psi_3^{+-})\}\rangle$\\$+2(\Phi^A-\Phi^V)\langle\psi^Lr_V\{\Psi_2(x_3+y-1)[2(1-x_1')r_{\Sigma}+x_3-1]$\\$+\Psi_4[x_3-x_2'r_{\Sigma}(x_3+y-1)-x_3'(2x_3r_{\Sigma}+yr_{\Sigma}-2r_{\Sigma}-1)]\}$\\$-r_c\psi^t[\Psi_4(1-x_1')r_{\Sigma}+(x_3-1)\Psi_2]\rangle$}\\\\
		
		$H_{B_{a2}}^{SP}$
		&\tabincell{c}{$\Psi_3^{-+}\Phi^T\{\psi^Lr_V(x_3+y-1)[(x_3-x_1')r_{\Sigma}-x_3+1]$\\$-(x_3-1)r_c(r_{\Sigma}-1)\psi^t\}$\\$+(\Phi^A+\Phi^V)\{r_c\psi^t[\Psi_4(1-x_1)(r_{\Sigma}-1)-(x_3-1)\Psi_2r_{\Sigma}]$\\$+\psi^Lr_V(x_3+y-1)[\Psi_4(1-x_1')(r_{\Sigma}-1)+(x_3-1)\Psi_2r_{\Sigma}]\}$}
		&\tabincell{c}{$-\Psi_3^{-+}\Phi^T\langle\psi^Lr_V\{x_3'[r_{\Sigma}(x_3+2y-1)+4(x_3+y-1)]$\\$ +2(x_3+y-1)[x_2'(r_{\Sigma}+2)+x_3-1]\}-2(x_3-1)r_c\psi^t\rangle$\\$+2(\Phi^A+\Phi^V)\langle\psi^Lr_V\{\Psi_2(x_3-2x_1'+1)r_{\Sigma}(x_3+y-1)$\\$-\Psi_4[(x_3-1)(r_{\Sigma}x_3'+2x_3'+x_2')+y(1-x_1')]\}-r_c\psi^t\{\Psi_4(1-x_1')+(x_3-1)\Psi_2r_{\Sigma}\}\rangle$}\\\\
		
		$H_{B_{a3}}^{LL}$
		&\tabincell{c}{$\frac{3}{2}x_3'r_{\Sigma}\Phi^T[\psi^Lr_V(x_3+y-1)(\Psi_3^{-+}-\Psi_3^{+-})-r_c\psi^t(\Psi_3^{-+}+\Psi_3^{+-})]$\\$-x_1r_{\Sigma}(\Phi^A-\Phi^V)\{\Psi_2r_c\psi^t+\psi^Lr_V[\Psi_4x_3'-\Psi_2(x_3+y-1)]\}$}
		&\tabincell{c}{$3x_3'r_{\Sigma}\Phi^T\{\psi^Lr_V[\Psi_3^{-+}(x_3+y-1)-\Psi_3^{+-}(1+x_1-x_3-y)]-r_c\psi^t(\Psi_3^{-+}+\Psi_3^{+-})\}$\\$+2r_{\Sigma}(\Phi^V-\Phi^A)\{[(x_3+y-1)\Psi_3^{-+}+(1+x_1-x_3+y)\Psi_3^{+-}\Phi^L]-r_c(\Psi_3^{-+}+\Psi_3^{+-})\Psi^t\}$}\\\\
		
		$H_{B_{a3}}^{SP}$
		&\tabincell{c}{$\frac{1}{2}x_1\psi^Lx_3'r_{\Sigma}r_V\Phi^T(\Psi_3^{-+}-\Psi_3^{+-})$\\$+x_1(1-r_{\Sigma})(\Phi^A+\Phi^V)\{\psi^Lr_V[\Psi_2(x_3+y-1)-\Psi_4x_3']-\Psi_2r_c\psi^t\}$}
		&\tabincell{c}{$x_1x_3'r_V\Phi^T\psi^L(r_{\Sigma}\Psi_3 ^{-+}+2\Psi_3 ^{+-})$\\$+2(\Phi^A+\Phi^V)\{\psi^Lr_V[\Psi_4x_3'(x_3'r_{\Sigma}-x_1)-\Psi_2(x_3+y-1)(2x_3'r_{\Sigma}-x_1)]-x_1\Psi_2r_c\psi^t\}$}\\\\
		
		$H_{B_{a4}}^{LL}$
    	&\tabincell{c}{$\frac{1}{2}\Psi_3^{+-}r_{\Sigma}\Phi^T\{(x_2'-1)[r_V\psi^L(1-x_3-y)-r_c\psi^t]+(x_3-1)\psi^Lx_3'r_V\}$\\$+(x_3-1)(\Phi^A-\Phi^V)\{\psi^Lr_V[\Psi_2(r_{\Sigma}-1)(x_3+y-1)-\Psi_4x_3'r_{\Sigma}]$\\$-\Psi_2r_c(r_{\Sigma}-1)\psi^t\}$}
    	&\tabincell{c}{$r_{\Sigma}\Phi^T\{\psi^Lr_V[(x_3-1)x_3'(\Psi_3^{-+}+\Psi_3^{+-})$\\$+(x_2'-1)(x_3+y-1)(2\Psi_3^{-+}-\Psi_3^{+-})]-r_c(x_2'-1)\Psi_3^{+-}\psi^t\}$\\$+2(x_3-1)(\Phi^A-\Phi^V)\{\Psi_2r_c\psi^t+\psi^Lr_V[\Psi_4x_3'-\Psi_2(x_3+y-1)]\}$}\\\\	
		
		$H_{B_{a4}}^{SP}$
		&\tabincell{c}{$(r_{\Sigma}-1)(\Phi^A+\Phi^V)\{\psi^Lr_V(x_3+y-1)$\\$[(x_3-1)\Psi_2-\Psi_4(x_2'-1)]-r_c\psi^t[\Psi_4(x_2'-1)+(x_3-1)\Psi_2]\}$\\$-\frac{1}{2}r_{\Sigma}\Phi^T\{r_c\Psi_3^{-+}(x_2'-1)\psi^t$\\$+\psi^Lr_V[\Psi_3^{-+}(x_2'-1)(x_3+y-1)-(x_3-1)x_3'\Psi_3^{+-}]\}$}
		&\tabincell{c}{$2(\Phi^A+\Phi^V)\{r_c\psi^t[\Psi_4(x_2'-1)+(x_3-1)\Psi_2]$\\$+\psi^Lr_V[\Psi_4(x_2'-1)(x_3-x_3'r_{\Sigma}+y-1)-(x_3-1)\Psi_2(x_3+y-1)]\}$\\$+\Phi^T\{\psi^Lr_V[\Psi_3^{-+}(x_2'-1)(r_{\Sigma}+4)(x_3+y-1)$\\$-2(x_3-1)x_3'\Psi_3^{+-}]-r_c\Psi_3^{-+}(x_2'-1)r_{\Sigma}\psi^t\}$}\\		
		\hline\hline
	\end{tabular}
\end{table}

\end{appendix}


\begin{thebibliography}{99}

\bibitem{Buras:2003yc}
A.~J.~Buras, R.~Fleischer, S.~Recksiegel, and F.~Schwab,
The $B \to \pi K$ puzzle and its relation to rare B and K decays,
Eur. Phys. J. C \textbf{32}, 45 (2003).

\bibitem{Baek:2004rp}
S.~Baek, P.~Hamel, D.~London, A.~Datta, and D.~A.~Suprun,
The $B \to \pi K$ puzzle and new physics,
Phys. Rev. D \textbf{71}, 057502 (2005).

\bibitem{Buras:2003dj}
A.~J.~Buras, R.~Fleischer, S.~Recksiegel, and F.~Schwab,
$B \to \pi \pi$, New Physics in $B \to \pi K$ and Implications for Rare K and B Decays,
Phys. Rev. Lett. \textbf{92}, 101804 (2004).

\bibitem{Buras:2004ub}
A.~J.~Buras, R.~Fleischer, S.~Recksiegel, and F.~Schwab,
Anatomy of prominent B and K decays and signatures of CP violating new physics in the electroweak penguin sector,
Nucl. Phys.  \textbf{B697}, 133 (2004).

\bibitem{Gronau:1998ep}
M.~Gronau and J.~L.~Rosner,
Combining CP asymmetries in $B \to K \pi$ decays,
Phys. Rev. D \textbf{59}, 113002 (1999).

\bibitem{LHCb:2020dpr}
R.~Aaij \textit{et al.} (LHCb Collaboration),
Measurement of CP Violation in the Decay $B^{+} \to K^{+} \pi^{0}$,
Phys. Rev. Lett. \textbf{126}, 091802 (2021).

\bibitem{Baek:2007yy}
S.~Baek and D.~London,
Is there still a $B \to \pi K$ puzzle,
Phys. Lett. B \textbf{653}, 249 (2007).

\bibitem{Baek:2009pa}
S.~Baek, C.~W.~Chiang, and D.~London,
The $B \to \pi K$ Puzzle: 2009 update,
Phys. Lett. B \textbf{675}, 59 (2009).

\bibitem{Gronau:2005kz}
M.~Gronau,
A precise sum rule among four $B \to K \pi$ CP asymmetries,
Phys. Lett. B \textbf{627}, 82 (2005).

\bibitem{Fleischer:2017vrb}
R.~Fleischer, R.~Jaarsma, and K.~K.~Vos,
Towards new frontiers with $B\to\pi K$ decays,
Phys. Lett. B \textbf{785}, 525 (2018).

\bibitem{Li:2005kt}
H.~n.~Li, S.~Mishima, and A.~I.~Sanda,
Resolution to the $B \to \pi K$ puzzle,
Phys. Rev. D \textbf{72}, 114005 (2005).

\bibitem{Grossman:1999av}
Y.~Grossman, M.~Neubert, and A.~L.~Kagan,
Trojan penguins and isospin violation in hadronic B decays,
J. High Engry Phys. 10 (1999) 029.

\bibitem{Yoshikawa:2003hb}
T.~Yoshikawa,
A possibility of large electroweak penguin contribution in $B \to K \pi$ modes,
Phys. Rev. D \textbf{68}, 054023 (2003).

\bibitem{Mishima:2004um}
S.~Mishima and T.~Yoshikawa,
Large electroweak penguin contribution in $B \to K \pi$ and $\pi \pi$ decay modes,
Phys. Rev. D \textbf{70}, 094024 (2004).

\bibitem{Gronau:2003kj}
M.~Gronau and J.~L.~Rosner,
Rates and asymmetries in $B \to K \pi$ decays,
Phys. Lett. B \textbf{572}, 43 (2003).

\bibitem{Kim:2007kx}
C.~S.~Kim, S.~Oh, and Y.~W.~Yoon,
Analytic resolution of puzzle in $B \to K \pi$ decays,
Phys. Lett. B \textbf{665}, 231 (2008).

\bibitem{Fleischer:1994rs}
R.~Fleischer,
Search for the angle gamma in the electroweak penguin dominated decay $B_s \to \pi^0 \Phi$,
Phys. Lett. B \textbf{332}, 419 (1994).

\bibitem{Deshpande:1994yd}
N.~G.~Deshpande, X.~G.~He, and J.~Trampetic,
Unique signature of electroweak penguin in pure hadronic $B$ decays,
Phys. Lett. B \textbf{345}, 547 (1995).

\bibitem{Chen:1998dta}
Y.~H.~Chen, H.~Y.~Cheng, and B.~Tseng,
Charmless hadronic two-body decays of $B_s$ mesons,
Phys. Rev. D \textbf{59}, 074003 (1999).

\bibitem{Beneke:2003zv}
M.~Beneke and M.~Neubert,
QCD factorization for $B \to PP$ and $B \to PV$ decays,
Nucl. Phys. \textbf{B675}, 333 (2003).

\bibitem{Beneke:2006hg}
M.~Beneke, J.~Rohrer and D.~Yang,
Branching fractions, polarisation and asymmetries of $B \to VV$ decays,
Nucl. Phys. \textbf{B774}, 64 (2007).

\bibitem{Ali:2007ff}
A.~Ali, G.~Kramer, Y.~Li, C.~D.~Lu, Y.~L.~Shen, W.~Wang, and Y.~M.~Wang,
Charmless non-leptonic $B_s$ decays to $PP$, $PV$ and $VV$ final states in the pQCD approach,
Phys. Rev. D \textbf{76}, 074018 (2007).

\bibitem{Zou:2015iwa}
Z.~T.~Zou, A.~Ali, C.~D.~Lu, X.~Liu and Y.~Li,
Improved estimates of the $B_s \to VV$ decays in perturbative QCD approach,
Phys. Rev. D \textbf{91}, 054033 (2015).

\bibitem{Chang:2015wba}
Q.~Chang, X.~Hu, J.~Sun, and Y.~Yang,
Probing spectator scattering and annihilation corrections in $B_s \to PV$ decays,
Phys. Rev. D \textbf{91}, 074026 (2015).

\bibitem{Wang:2008rk}
W.~Wang, Y.~M.~Wang, D.~S.~Yang and C.~D.~Lu,
Charmless two-body $B_s\to VP$ decays in soft-collinear-effective-theory,
Phys. Rev. D \textbf{78}, 034011 (2008).

\bibitem{Yan:2017nlj}
D.~C.~Yan, P.~Yang, X.~Liu, and Z.~J.~Xiao,
Anatomy of $B_s \to PV $ decays and effects of next-to-leading order contributions in the perturbative QCD factorization approach,
Nucl. Phys. \textbf{B931}, 79 (2018).


\bibitem{Cheng:2009mu}
H.~Y.~Cheng and C.~K.~Chua,
{QCD} factorization for charmless hadronic $B_s$ decays revisited,
Phys. Rev. D \textbf{80}, 114026 (2009).

\bibitem{Wang:2017rmh}
C.~Wang, S.~H.~Zhou, Y.~Li, and C.~D.~Lu,
Global analysis of charmless $B$ decays into two vector mesons in soft-collinear effective theory,
Phys. Rev. D \textbf{96}, 073004 (2017).

\bibitem{Faisel:2011kq}
G.~Faisel,
Supersymmetric contributions to $\bar{B}_s \to \phi \pi^0$ and $\bar{B}_s \to \phi \rho^0$ decays in SCET,
J. High Engry Phys. 08 (2012) 031.

\bibitem{Faisel:2013nra}
G.~Faisel,
Charged Higgs contribution to $\bar{B}_s \rightarrow \phi \pi^0 $ and $\bar{B}_s \rightarrow \phi \rho^0 $,
Phys. Lett. B \textbf{731}, 279 (2014).

\bibitem{Faisel:2017glo}
G.~Faisel and J.~Tandean,
Connecting $ b\to s\ell \bar{\ell} $ anomalies to enhanced rare nonleptonic $ \bar{B}_s^0 $ decays in $Z'$ model,
J. High Engry Phys. 02 (2018) 074.

\bibitem{Hofer:2010ee}
L.~Hofer, D.~Scherer, and L.~Vernazza,
$B_s \to \phi \rho^0$ and $B_s \to \phi \pi^0$ as a handle on isospin-violating new physics,
J. High Engry Phys. 02 (2011) 080.

\bibitem{Hofer:2011yg}
L.~Hofer, D.~Scherer, and L.~Vernazza,
Probing new physics in electroweak penguins through $B_d$ and $B_s$ decays,
J. Phys. Conf. Ser. \textbf{335}, 012039 (2011).

\bibitem{Faisel:2014dna}
G.~Faisel,
New physics contributions to $\bar{B}_s \to \pi^0(\rho^0 )\,\eta^{(')} $ decays,
Eur. Phys. J. C \textbf{77}, 380 (2017).

\bibitem{LHCb:2016vqn}
R.~Aaij \textit{et al.} (LHCb Collaboration),
Observation of the decay $B^0_s \to \phi\pi^+\pi^-$ and evidence for $B^0 \to \phi\pi^+\pi^-$,
Phys. Rev. D \textbf{95}, 012006 (2017).

\bibitem{Ecker:2000zr}
G.~Ecker, G.~Isidori, G.~Muller, H.~Neufeld, and A.~Pich,
Electromagnetism in nonleptonic weak interactions,
Nucl. Phys. \textbf{B591}, 419 (2000).

\bibitem{Dery:2020lbc}
A.~Dery, M.~Ghosh, Y.~Grossman, and S.~Schacht,
SU(3)$_{F}$ analysis for beauty baryon decays,
J. High Energy Phys. 03 (2020) 165.

\bibitem{Keum:2000ms}
Y.~Y.~Keum and H.~n.~Li,
Nonleptonic charmless B decays: Factorization versus perturbative QCD,
Phys. Rev. D \textbf{63}, 074006 (2001).

\bibitem{Li:1994cka}
H.~n.~Li and H.~L.~Yu,
Extraction of $V_{ub}$ from decay $B\rightarrow \pi l \nu$,
Phys. Rev. Lett. \textbf{74}, 4388 (1995).

\bibitem{Kurimoto:2001zj}
T.~Kurimoto, H.~n.~Li, and A.~I.~Sanda,
Leading power contributions to $B\rightarrow \pi, \rho$ transition form-factors,
Phys. Rev. D \textbf{65}, 014007 (2001).

\bibitem{Lu:2002ny}
C.~D.~Lu and M.~Z.~Yang,
B to light meson transition form-factors calculated in perturbative QCD approach,
Eur. Phys. J. C \textbf{28}, 515  (2003).


\bibitem{Wang:2010ni}
W.~Wang,
B to tensor meson form factors in the perturbative QCD approach,
Phys. Rev. D \textbf{83}, 014008 (2011).

\bibitem{Rui:2021kbn}
Z.~Rui, Y.~Li, and H.~n.~Li,
Four-body decays $B_{(s)} \rightarrow (K\pi)_{S/P} (K\pi)_{S/P}$ in the perturbative QCD approach,
J. High Energy Phys. 05 (2021) 082.

\bibitem{Chai:2022ptk}
J.~Chai, S.~Cheng, Y.~h.~Ju, D.~C.~Yan, C.~D.~L\"u, and Z.~J.~Xiao,
Charmless two-body $B$ meson decays in the perturbative QCD factorization approach,
Chin. Phys. C \textbf{46}, 123103 (2022).

\bibitem{prd59094014}
H.~H.~Shih, S.~C.~Lee, and H.~n.~Li,
The $\Lambda_b \to p l \bar{\nu}$ decay in perturbative QCD,
Phys. Rev. D \textbf{59}, 094014 (1999).

\bibitem{prd61114002}
H.~H.~Shih, S.~C.~Lee, and H.~n.~Li,
Applicability of perturbative QCD to  $\Lambda_b \to \Lambda_c$  decays,
Phys. Rev. D \textbf{61}, 114002 (2000).

\bibitem{cjp39328}
H.~H.~Shih, S.~C.~Lee, and H.~N.~Li,
Asymmetry parameter in the polarized $\Lambda_b \to \Lambda_c l \bar{\nu}$ decay,
Chin. J. Phys. \textbf{39}, 328 (2001).

\bibitem{prd65074030}
C.~H.~Chou, H.~H.~Shih, S.~C.~Lee, and H.~n.~Li,
$\Lambda_b \to \Lambda J/\psi$ decay in perturbative QCD,
Phys. Rev. D \textbf{65}, 074030 (2002).

\bibitem{prd74034026}
X.~G.~He, T.~Li, X.~Q.~Li, and Y.~M.~Wang,
PQCD calculation for $\Lambda_b \to \Lambda \gamma$ in the standard model,
Phys. Rev. D \textbf{74}, 034026 (2006).

\bibitem{prd80034011}
C.~D.~Lu, Y.~M.~Wang, H.~Zou, A.~Ali, and G.~Kramer,
Anatomy of the pQCD approach to the baryonic decays $\Lambda_b\to p\pi, pK$,
Phys. Rev. D \textbf{80}, 034011 (2009).

\bibitem{220204804}
J.~J.~Han, Y.~Li, H.~n.~Li, Y.~L.~Shen, Z.~J.~Xiao, and F.~S.~Yu,
$\Lambda_b\to p$ transition form factors in perturbative QCD,
Eur. Phys. J. C \textbf{82}, 686 (2022).

\bibitem{220209181}
C.~Q.~Zhang, J.~M.~Li, M.~K.~Jia, and Z.~Rui,
Nonleptonic two-body decays of $\Lambda_b\to \Lambda_c \pi, \Lambda_cK$ in the perturbative QCD approach,
Phys. Rev. D \textbf{105}, 073005 (2022).

\bibitem{prd106053005}
Z.~Rui, C.~Q.~Zhang, J.~M.~Li, and M.~K.~Jia,
Investigating the color-suppressed decays $\Lambda_b\to \Lambda \psi$ in the perturbative QCD approach,
Phys. Rev. D \textbf{106}, 053005 (2022).

\bibitem{221015357}
Z.~Rui, J.~M.~Li, and C.~Q.~Zhang,
Estimates of exchange topological contributions and $CP$-violating observables in $\Lambda_b\to \Lambda \phi$ decay,
Phys. Rev. D \textbf{107}, 053009 (2023).


\bibitem{230213785}
Z.~Rui, J.~M.~Li, and C.~Q.~Zhang,
Mixing effects of $\eta-\eta'$ in $\Lambda_b\rightarrow \Lambda \eta^{(')}$ decays,
Phys. Rev. D \textbf{107}, 093008 (2023).

\bibitem{Ali:2012zza}
A.~Ali, C.~Hambrock, and A.~Y.~Parkhomenko,
Light-cone wave functions of heavy baryons,
Theor. Math. Phys. \textbf{170}, 2 (2012).

\bibitem{plb665197}
P.~Ball, V.~M.~Braun, and E.~Gardi,
Distribution amplitudes of the $\Lambda_b$ baryon in QCD,
Phys. Lett. B \textbf{665}, 197 (2008).

\bibitem{jhep112013191}
G.~Bell, T.~Feldmann, Y.~M.~Wang, and M.~W.~Y.~Yip,
Light-cone distribution amplitudes for heavy-quark hadrons,
J. High Energy Phys. 11 (2013) 191.

\bibitem{jhep022016179}
Y.~M.~Wang and Y.~L.~Shen,
Perturbative corrections to $\Lambda_b \to \Lambda$ form factors from QCD Light-cone sum rules,
J. High Energy Phys. 02 (2016) 179.

\bibitem{epjc732302}
A.~Ali, C.~Hambrock, A.~Y.~Parkhomenko, and W.~Wang,
Light-cone distribution amplitudes of the ground state bottom baryons in HQET,
Eur. Phys. J. C \textbf{73}, 2302 (2013).

\bibitem{plb738334}
V.~M.~Braun, S.~E.~Derkachov, and A.~N.~Manashov,
Integrability of the evolution equations for heavy-light baryon distribution amplitudes,
Phys. Lett. B \textbf{738}, 334 (2014).

\bibitem{zpc42569}
V.~L.~Chernyak, A.~A.~Ogloblin, and I.~R.~Zhitnitsky,
Wave functions of octet baryons,
Z. Phys. C \textbf{42}, 569 (1989).


\bibitem{Farrar:1988vz}
G.~R.~Farrar, H.~Zhang, A.~A.~Ogloblin, and I.~R.~Zhitnitsky,
Baryon wave functions and cross-sections for photon annihilation to baryon pairs,
Nucl. Phys. \textbf{B311}, 585 (1989).

\bibitem{Liu:2014uha}
Y.~L.~Liu, C.~Y.~Cui, and M.~Q.~Huang,
Higher order light-cone distribution amplitudes of the $\Lambda$ baryon,
Eur. Phys. J. C \textbf{74}, 3041 (2014).

\bibitem{Liu:2008yg}
Y.~L.~Liu and M.~Q.~Huang,
Distribution amplitudes of $\Sigma$ and $\Lambda$ and their electromagnetic form factors,
Nucl. Phys. \textbf{A821}, 80 (2009).

\bibitem{jhep020702016}
G.~S.~Bali, V.~M.~Braun, M.~G\"ockeler, M.~Gruber, F.~Hutzler, A.~Sch\"afer, R.~W.~Schiel, J.~Simeth, W.~S\"oldner, A.~Sternbeck \textit{et al.},
Light-cone distribution amplitudes of the baryon octet,
J. High Energy Phys. 02 (2016) 070.

\bibitem{prd89094511}
V.~M.~Braun, S.~Collins, B.~Gl\"a\ss{}le, M.~G\"ockeler, A.~Sch\"afer, R.~W.~Schiel, W.~S\"oldner, A.~Sternbeck, and P.~Wein,
Light-cone distribution amplitudes of the nucleon and negative parity nucleon resonances from lattice QCD,
Phys. Rev. D \textbf{89}, 094511 (2014).

\bibitem{epja55116}
G.~S.~Bali \textit{et al.} (RQCD Collaboration),
Light-cone distribution amplitudes of octet baryons from lattice QCD,
Eur. Phys. J. A \textbf{55}, 116 (2019).

\bibitem{Deng:2023csv}
Z.~F.~Deng, C.~Han, W.~Wang, J.~Zeng, and J.~L.~Zhang,
Light-cone distribution amplitudes of a light baryon in large-momentum effective theory,
J. High Energy Phys. 07 (2023) 191.

\bibitem{prd71114008}
C.~H.~Chen and H.~N.~Li,
Nonfactorizable contributions to $B$ meson decays into charmonia,
Phys. Rev. D \textbf{71}, 114008 (2005).

\bibitem{Sun:2008ew}
J.~F.~Sun, D.~S.~Du, and Y.~L.~Yang,
Study of $B_c \to J/\psi \pi$, $\eta_c \pi$ decays with perturbative QCD approach,
Eur. Phys. J. C \textbf{60}, 107 (2009).

\bibitem{prd90114030}
Z.~Rui and Z.~T.~Zou,
$S$-wave ground state charmonium decays of $B_c$ mesons in the perturbative QCD approach,
Phys. Rev. D \textbf{90}, 114030 (2014).

\bibitem{epjc76564}
Z.~Rui, H.~Li, G.~x.~Wang, and Y.~Xiao,
Semileptonic decays of $B_c$ meson to S-wave charmonium states in the perturbative QCD approach,
Eur. Phys. J. C \textbf{76},  564 (2016).


\bibitem{Ball:1998sk}
P.~Ball, V.~M.~Braun, Y.~Koike, and K.~Tanaka,
Higher twist distribution amplitudes of vector mesons in QCD: Formalism and twist-three distributions,
Nucl. Phys. \textbf{B529}, 323 (1998).

\bibitem{Ball:1998ff}
P.~Ball and V.~M.~Braun,
Higher twist distribution amplitudes of vector mesons in QCD: Twist - 4 distributions and meson mass corrections,
Nucl. Phys. \textbf{B543}, 201 (1999).

\bibitem{Ball:2007rt}
P.~Ball and G.~W.~Jones,
Twist-3 distribution amplitudes of $K^*$ and phi mesons,
J. High Energy Phys. 03 (2007) 069.

\bibitem{Rui:2017fje}
Z.~Rui, Y.~Li, and Z.~J.~Xiao,
Branching ratios, $CP$ asymmetries and polarizations of $B\to \psi(2S) V$ decays,
Eur. Phys. J. C \textbf{77}, 610 (2017).

\bibitem{Buchalla:1995vs}
G.~Buchalla, A.~J.~Buras, and M.~E.~Lautenbacher,
Weak decays beyond leading logarithms,
Rev. Mod. Phys. \textbf{68}, 1125 (1996).

\bibitem{zpc55659}
J.~G.~Korner and M.~Kramer,
Exclusive nonleptonic charm baryon decays,
Z. Phys. C \textbf{55}, 659 (1992).

\bibitem{prd562799}
H.~Y.~Cheng,
Nonleptonic weak decays of bottom baryons,
Phys. Rev. D \textbf{56}, 2799 (1997); \textbf{99}, 079901(E) (2019).

\bibitem{plb72427}
R.~Aaij \textit{et al.} (LHCb Collaboration),
Measurements of the $\Lambda_b^0 \to J/\psi \Lambda$ decay amplitudes and the $\Lambda_b^0$ polarisation in $pp$ collisions at $\sqrt{s} = 7$ TeV,
Phys. Lett. B \textbf{724}, 27 (2013).

\bibitem{Gutsche:2013oea}
T.~Gutsche, M.~A.~Ivanov, J.~G.~K\"orner, V.~E.~Lyubovitskij, and P.~Santorelli,
Polarization effects in the cascade decay $\Lambda_b\to \Lambda (\to p \pi) + J/\psi (\to l^+l^-)$ in the covariant confined quark model,
Phys. Rev. D \textbf{88}, 114018 (2013).


\bibitem{Dalitz:1964es}
R.~H.~Dalitz and F.~Von Hippel,
Electromagnetic $\Lambda-\sigma^0$ mixing and charge symmetry for the $\Lambda$-hyperon,
Phys. Lett. \textbf{10}, 153 (1964).

\bibitem{Gal:1967oxu}
A.~Gal and F.~Scheck,
Electromagnetic mass splittings of mesons and baryons in the quark model,
Nucl. Phys. \textbf{B2}, 110 (1967).

\bibitem{Isgur:1979ed}
N.~Isgur,
Isospin violating mass differences and mixing angles: The role of quark masses,
Phys. Rev. D \textbf{21}, 779 (1980); \textbf{23}, 817(E) (1981).

\bibitem{Gasser:1982ap}
J.~Gasser and H.~Leutwyler,
Quark masses,
Phys. Rep. \textbf{87}, 77 (1982).

\bibitem{Yagisawa:2001gz}
N.~Yagisawa, T.~Hatsuda, and A.~Hayashigaki,
In-medium $\Sigma^0-\Lambda$ mixing in QCD sum rules,
Nucl. Phys. \textbf{A699}, 665 (2002).

\bibitem{Zhu:1998ai}
S.~L.~Zhu, W.~Y.~P.~Hwang, and Z.~s.~Yang,
The possible $\Sigma^0-\Lambda$ mixing in QCD sum rule,
Phys. Rev. D \textbf{57}, 1524 (1998).

\bibitem{Radici:2001pq}
M.~Radici,
T odd fragmentation functions,
Nucl. Phys. \textbf{A699}, 144 (2002).

\bibitem{Horsley:2014koa}
R.~Horsley, J.~Najjar, Y.~Nakamura, H.~Perlt, D.~Pleiter, P.~E.~L.~Rakow, G.~Schierholz, A.~Schiller, H.~St\"uben and J.~M.~Zanotti,
Lattice determination of Sigma - Lambda mixing,
Phys. Rev. D \textbf{91}, 074512 (2015).

\bibitem{Aliev:2015ela}
T.~M.~Aliev, T.~Barakat, and M.~Savc\i{},
Determination of the $\Sigma-\Lambda$ mixing angle from QCD sum rules,
J. High Energy Phys. 08 (2016) 068.

\bibitem{Kordov:2019oer}
Z.~R.~Kordov \textit{et al.} (CSSM/QCDSF/UKQCD Collaborations),
Electromagnetic contribution to $\Sigma-\Lambda$ mixing using lattice QCD+QED,
Phys. Rev. D \textbf{101}, 034517 (2020).


\bibitem{Geng:2020tlx}
C.~Q.~Geng, C.~W.~Liu, and T.~H.~Tsai,
Mixing effects of $\Sigma^0-\Lambda^0$ in $\Lambda_c^+$ decays,
Phys. Rev. D \textbf{101},  054005 (2020).

\bibitem{Franklin:2020bvb}
J.~Franklin,
Quark model relations for b-baryon decay,
J. Phys. G \textbf{47}, 085001 (2020).

\bibitem{Franklin:1981rc}
J.~Franklin, D.~B.~Lichtenberg, W.~Namgung, and D.~Carydas,
Wave function mixing of flavor degenerate baryons,
Phys. Rev. D \textbf{24}, 2910 (1981).

\bibitem{pdg2022}
R.L. Workman \textit{et al.} (Particle Data Group), Prog. Theor. Exp. Phys. \textbf{2022}, 083C01 (2022).

\bibitem{Lu:2000em}
C.~D.~Lu, K.~Ukai, and M.~Z.~Yang,
Branching ratio and CP violation of $B \to \pi \pi$ decays in perturbative QCD approach,
Phys. Rev. D \textbf{63}, 074009 (2001).

\bibitem{LHCb:2019aci}
R.~Aaij \textit{et al.} (LHCb Collaboration),
Isospin Amplitudes in $\Lambda_b^0\to J/\psi \Lambda(\Sigma^0)$ and $\Xi_b^0\to J/\psi \Xi^0(\Lambda)$ decays,
Phys. Rev. Lett. \textbf{124}, 111802 (2020).





\end{thebibliography}
\end{document}